\documentclass[10pt,journal,compsoc]{IEEEtran}

 \usepackage{cite}

\usepackage{graphicx}
\usepackage{multirow}
\usepackage{hhline}
\usepackage{comment}
\usepackage{bigstrut}
\usepackage[dvipsnames]{xcolor}
\usepackage[colorlinks=false, linkcolor=blue, citecolor=blue, bookmarks=true,pagebackref=false]{hyperref}
\usepackage{hyphenat}
\usepackage{subcaption}
\usepackage{float}
\usepackage{fancybox}
\usepackage[flushleft]{threeparttable}
\usepackage{array,booktabs}
\usepackage{amsmath}
\usepackage{outlines}
\usepackage[linesnumbered,ruled]{algorithm2e}
\usepackage[misc]{ifsym}
\usepackage{arydshln}
\usepackage{balance}
\usepackage{amsmath}%
\usepackage{MnSymbol}%
\usepackage{wasysym}%
\usepackage{enumitem}
\usepackage{fullpage}
\usepackage{times}
\usepackage{fancyhdr}

\usepackage{array}
\newcolumntype{R}[1]{>{\raggedleft\let\newline\\\arraybackslash\hspace{2pt}}m{#1}}

\newcommand{\rqbox}[1]{
\begin{center}
\vspace{-0.5cm}
\cornersize{.2}
\setlength{\fboxsep}{7pt}
\ovalbox{\begin{minipage}{3in}
{\em #1}
\end{minipage}}
\vspace{0cm}

\end{center}}

\renewcommand{\IEEEbibitemsep}{0pt plus 0.3pt}
\makeatletter
\IEEEtriggercmd{\reset@font\normalfont\fontsize{7pt}{7pt}\selectfont}
\makeatother
\IEEEtriggeratref{1}

\newcommand{\etal}[1]{${\textit{et al.}}$}

\newcolumntype{M}[1]{>{\centering\arraybackslash}m{#1}}

\ifCLASSINFOpdf
\else
\fi
\begin{document}
\bstctlcite{IEEEexample:BSTcontrol}
%
\title{Impact of Discretization Noise of the Dependent variable on Machine Learning Classifiers in Software Engineering}

%
%
%
%

\author{Gopi~Krishnan~Rajbahadur,
Shaowei~Wang,
Yasutaka~Kamei, and
Ahmed~E.~Hassan
\IEEEcompsocitemizethanks{\IEEEcompsocthanksitem Gopi~Krishnan~Rajbahadur, Shaowei~Wang and Ahmed~E.~Hassan are with Software Analysis and Intelligence Lab (SAIL), School of Computing, Queen's University, Canada.\protect\\
E-mail: $\{$krishnan, shaowei, ahmed$\}$@cs.queensu.ca
\IEEEcompsocthanksitem Yasutaka~Kamei is with Principles of Software Languages (POSL) Lab, Graduate School and Faulty of Information Science and Electrical Engineering, Kyushu University, Japan.

Email: kamei@ait.kyushu-u.ac.jp

\IEEEcompsocthanksitem Shaowei Wang is the corresponding author

}

}

\IEEEtitleabstractindextext{%
\begin{abstract}
Researchers usually discretize a continuous dependent variable into two target classes by introducing an artificial discretization threshold (e.g., median). However, such discretization may introduce noise (i.e., discretization noise) due to ambiguous class loyalty of data points that are close to the artificial threshold. Previous studies do not provide a clear directive on the impact of discretization noise on the classifiers and how to handle such noise. In this paper, we propose a framework to help researchers and practitioners systematically estimate the impact of discretization noise on classifiers in terms of its impact on various performance measures and the interpretation of classifiers. Through a case study of 7 software engineering datasets, we find that: 1) discretization noise affects the different performance measures of a classifier differently for different datasets; 2) Though the interpretation of the classifiers are impacted by the discretization noise on the whole, the top 3 most important features are not affected by the discretization noise. Therefore, we suggest that practitioners and researchers use our framework to understand the impact of discretization noise on the performance of their built classifiers and estimate the exact amount of discretization noise to be discarded from the dataset to avoid the negative impact of such noise.

\end{abstract}

}

\maketitle

\IEEEdisplaynontitleabstractindextext

%
\IEEEpeerreviewmaketitle

\vspace{-2cm}

\section{Introduction}\label{sec:introduction}


\IEEEPARstart{M}achine learning classifiers are widely used throughout software engineering studies. Some of the most common uses of classifiers include predicting defects~\cite{rajbahadur2017impact,Ghotra:2015:RIC:2818754.2818850,krishna2017less,nam2017heterogeneous}, bug-fix times~\cite{jiang2013will}, understanding the features that impact the defect proneness of a software system~\cite{cataldo2009software,mcintosh2014impact,mockus2010organizational,menzies2007strangest}.

Usually, classifiers are trained on labeled data points and are used to predict the target class of the unlabeled data points. In the absence of pre-defined class labels and the availability of only the continuous dependent variable, researchers usually discretize\footnote{Discretization is the process of turning numerical data into discrete data with finite intervals~\cite{garcia2012survey}} the continuous \textbf{dependent} variable into artificial target classes. Such discretization might be based on domain knowledge~\cite{guo2010characterizing}, a phenomenon that the study wishes to observe~\cite{jiang2013will}, or in many cases when there are no imposed target classes, an artificial discretization threshold is used to discretize the target variable into binary (or n-ary) classes~\cite{de1998investigation,el2001modelling,Tian:2015,wang2017}.

However a plethora of prior studies note that the discretization of the continuous dependent variable could be detrimental to the performance of classifier and may produce misleading results~\cite{cohen1983cost,dawson2012dichotomizing,royston2006dichotomizing,decoster2009conceptual}. One alternative approach to avoid such discretization noise would be to train regression models on the continuous dependent variable and then discretize the predicted outcome afterwards~\cite{rajbahadur2017impact}. But as Rajbahadur~\textit{\etal}~\cite{rajbahadur2017impact} observe, only when there is a significant class imbalance in the dataset, classification through regression would yield better results. Therefore, the discretization of continuous dependent variable with artificial discretization thresholds is still widely practiced in software engineering as evidenced by~\cite{guo2010characterizing,gay2010automatically,jiang2013will,schumann2009software,jalali2008optimizing,wang2017,Tian:2015,hassan2018studying}.

The other problem with such a discretization approach is that the data points that are very close to the discretization threshold (e.g., median) get class labels that might not be reflective of the true class to which they belong. While many previous studies explore the harmful impacts of discretizing the continuous dependent variables on the performance of the classifiers~\cite{cohen1983cost,dawson2012dichotomizing,royston2006dichotomizing,decoster2009conceptual}, the problem of data points with ambiguous class labels and its impact on the classifiers is completely unexplored. For instance, consider the example of determining whether a bug was closed fast or slow. A domain-expert might decide that any bug closed within a week is a “fast-closed” bug. However, such a discretization rule for the dependent variable would lead to noise for data points close to that 7-days threshold. For instance, a bug that is closed within 7 days and 1 min would be considered as a “slow-closed” bug. Such discretization introduces noise in the data that is used for training the classifiers. We define such noise as the \textit{discretization noise} and the data points whose ambiguous class labels that generate the discretization noise as the \textit{noisy points}.

In summary, discretization of continuous dependent variable is both problematic and generates discretization noise. However, the practice of discretizing the continuous dependent variable still remains a widely used practice in software engineering without any consideration to the generated discretization noise.

Therefore the main goal of our study is two-fold: first, to introduce awareness among the software engineering researchers and practitioners about previously unexplored discretization noise. Second, we provide them with a framework - a systematic and rigorous method for exploring the impact of discretization noise on their classifier of choice for any given dataset.

We highlight the capability of our framework by conducting our study on four different types of software engineering datasets (Q\&A websites data (4 websites), Linux patch acceptance time data, bug-fix delay data, mobile app ratings data), as a binary classification problem, where the discretization noise is caused by the discretization of a continuous dependent variable around artificial discretization thresholds. We applied our proposed framework to four different families of classifiers (i.e., Random forest classifier (RFCM), Logistic regression (LR), Classification and Regression Trees (CART), and K-Nearest Neighbors (KNN)) to analyse the impact of discretization noise on the various common performance measures (i.e., \textit{Accuracy, Precision, Recall, Brier score, AUC, F-Measure, MCC}). In addition, we also analyse the impact of discretization noise on the interpretation of classifiers in terms of the derived feature importance ranks. The derived feature importance ranks is a rank list that reports the features of the dataset in the order of their influence on the classification.

We highlight our findings and suggestions as follows:
\begin{enumerate}[leftmargin=*]
  \item \textbf{The impact of discretization noise is inconsistent across multiple performance measures for different datasets across all the studied classifiers.} Though the impact on Recall is the most pronounced (up to 139\%), other performance measures - Precision, Brier score, F-Measure, and MCC are also impacted at least up to 43.19\% due to the inclusion or exclusion of discretization noise (both positively and negatively). \textbf{Therefore, we urge the researchers and practitioners to use our framework to analyse if (and how much) discretization noise exists and how to address it.}
  \item \textbf{Though the overall derived feature importance ranks are impacted by discretization noise, the importance ranks of the top three important features are not affected.} Therefore, in absence of any impact on the interpretation of the top \textit{x} features detected by our framework, one could include or discard the discretization noise in their datasets as recommended by our framework. Especially without being worried about the discretization noise's impact on the interpretation of the classifier.
\end{enumerate}

Finally, we also provide the framework as an R package (and guidelines for using our framework) to provide automated support to others who wish to revisit their prior results or to consider discretization in the future studies.


\textbf{Paper organization.} Section~\ref{sec:related} places our contribution relative to prior work. Section~\ref{sec:framework} outlines our framework. Section~\ref{sec:rqs} presents the impact of discretization noise on the performance of the classifiers and on the derived feature importance ranks of several software engineering datasets. Section~\ref{sec:disc} explores why discretization noise is sometimes detrimental and sometimes useful. In Section~\ref{sec:guideline} provide a user guideline for our framework. We then discuss the threats to the validity of our observations in Section~\ref{sec:threat}. Finally, we conclude our study in Section~\ref{sec:conclusion}.


\section{Related Work}\label{sec:related}
In this section, we discuss two groups of prior work: efforts that attempt to establish the impact of noise on classifiers and efforts that investigate problems with discretization on classifiers.
\subsection{Impact of Noise on Classifiers}
Discretization noise while being similar to class noise~\cite{zhu2004class}, differs from it because class noise is caused by data points with wrong class labels due to contradictory examples and random misclassifications~\cite{zhu2004class}. Whereas, we view discretization noise as data points with ambiguous class labels due to their proximity to the discretization threshold. While discretization noise is unexplored, a number of studies have explored the impact of noise on classifiers~\cite{zhu2004class,kim2011dealing,folleco2008software,khoshgoftaar2005enhancing,seiffert2014empirical}. For instance, Kim~\textit{\etal}~\cite{kim2011dealing} explored the impact of class noise on defect classifiers and provided insights on acceptable levels of noise in the data when building classifiers and the sensitivity of the various classifiers to class noise. Whereas Folleco~\etal~\cite{folleco2008software} explored the impact of class noise on various classifiers when using software quality data and found that the performance of the random forest classifiers is the most robust. Similarly, studies outside software engineering have also found class noise to be detrimental to the performance of a classifier~\cite{zhu2003eliminating,nettleton2010study}. On the contrary, Tantithamthavorn~\textit{\etal}~\cite{tantithamthavorn2015icse} demonstrated that the Precision of a classifier is not impacted by class noise and the most influential features (i.e., features in the top 3 ranks) remain the same irrespective of the amount of class noise that is present in the data. They suggested that the insights from the constructed classifiers can be used without worrying about the noise. However, they did note that filtering the noise increases the Recall of the classifiers.
Different from prior studies, we are not concerned with the impact of class noise and noise in the dataset that is generated at random~\cite{Harrell2001} or during data collection~\cite{mockus2008missing}. Rather, we provide a framework to analyse the impact of the discretization noise that is generated due to the discretization of the dependent variable (in the absence of labeled data) on the performance of classifiers and the derived feature importance ranks of a classifier in a systematic and repeatable manner.
\subsection{Impact of Discretization on Classifiers}

In software engineering, discretization has been used to transform both continuous independent and dependent variables into discrete classes~\cite{menzies2011inductive,rajbahadur2017impact,jiang2008can,ma2012transfer,menzies2010defect,zheng2015method}.

\begin{table}[htbp]
  \centering
\scriptsize
  \caption{Details of datasets used in the study}
  \begin{threeparttable}
    \begin{tabular}{p{1.85cm}|p{0.5cm}|p{1cm}|p{1cm}}
    \hline
    \textbf{Dataset} & \textbf{\#Size} & \textbf{\#Features} & \textbf{R(dependent variable)}\\
    \hline
    \textbf{Stack Overflow} & \multicolumn{1}{r|}{ 55,853} & \multicolumn{1}{r|}{ 28} & \multicolumn{1}{r}{ 0-9,981.40 mins} \\
    \textbf{Mathematics} & \multicolumn{1}{r|}{ 70,336} & \multicolumn{1}{r|}{ 27} & \multicolumn{1}{r}{ 0-30,073.72 mins} \\
    \textbf{Ask Ubuntu} & \multicolumn{1}{r|}{ 7,134} & \multicolumn{1}{r|}{ 26} & \multicolumn{1}{r}{ 0-31,638.55 mins} \\
    \textbf{Super User} & \multicolumn{1}{r|}{ 10,776} & \multicolumn{1}{r|}{ 27} & \multicolumn{1}{r}{ 0-51,376.33 mins} \\
    \textbf{Patch} & \multicolumn{1}{r|}{ 20,000} & \multicolumn{1}{r|}{ 22} & \multicolumn{1}{r}{ 0-1,266.92 days} \\
    \textbf{Bug-delay} & \multicolumn{1}{r|}{ 2,434} & \multicolumn{1}{r|}{ 23} & \multicolumn{1}{r}{-1,319.06*-1,990.27 days} \\
    \textbf{App-rating} & \multicolumn{1}{r|}{ 7,365} & \multicolumn{1}{r|}{ 22} & \multicolumn{1}{r}{ 1.41-4.97 stars} \\
    \hline
    \end{tabular}%
    \begin{tablenotes}
      \scriptsize
      \item *The dependent variable has negative value since some developers started fixing a bug before the bug was reported.
      \item R(x) - Range(x)
    \end{tablenotes}

\end{threeparttable}
  \label{tab:dataset_details}%
\end{table}%

\subsubsection{Discretization of the independent variable}
Yang~\textit{\etal}~\cite{yang2009discretization} and Garcia~\textit{\etal}~\cite{garcia2012survey} have shown that the discretization of the independent variables improves the performance of machine learning classifiers. True to that result, many studies in software engineering discretize the independent variables as part of data pre-processing before constructing classifiers~\cite{menzies2011inductive,ma2012transfer}. For instance, Jiang~\textit{\etal} investigated and found that while discretization of independent variables improves the performance of some classifiers, it did not universally benefit all classifiers~\cite{jiang2008can}. Contrary to those studies, Nam and Kim used a median based discretization of independent variables to assign class labels for un-labeled data points to train a cross-project defect prediction model~\cite{nam2015clami}. However, except for univariate discretization techniques, the discretization techniques used in the aforementioned studies cannot be used in our study, as they focus on discretizing an independent variable based on the information that is contained in the other independent variables or the class distribution of the dependent variable. But in our study, we focus on studying the impact of discretizing the dependent variable into meaningful target classes based on itself and domain knowledge. 
\subsubsection{Discretization of the dependent variable}
Mockus~\cite{mockus2008missing} suggested that many of the software engineering datasets have poor data quality which affects the insights that are offered by the software analytical models. This noise can arise from various sources in the case of labeled data. But when the class labels for the data are not available and a continuous dependent variable is discretized using a threshold to generate class labels~\cite{de1998investigation,el2001modelling,Tian:2015,zheng2015method}, we end up with discretization noise. But many researchers actively have argued against the practice of discretization. For instance, Altman~\textit{\etal}~\cite{altman2006cost} and Cohen~\cite{cohen1983cost} pointed out that discretization at the median of a continuous variable leads to a loss of information, and discretization at other cut points away from the center lead to a much greater information loss. In addition, several prior studies argued against the use of any data-driven cutpoint to discretize dependent variable as it introduced noise and bias~\cite{rucker2015researcher,dawson2012dichotomizing,royston2006dichotomizing}. MacCallum~\cite{maccallum2002practice} and  DeCoster~\textit{\etal}~\cite{decoster2009conceptual} further state that discretization is rarely ever justifiable and it is almost always safer not to discretize. However, several researchers also note that it is acceptable to discretize sometimes~\cite{decoster2011best,decoster2009conceptual,farrington2000some,kraemer2004categorical}. In summary, though there are some benefits associated with discretization, the majority of the research agrees that it is not a safe practice and should mostly be avoided. 
\subsubsection{Discretization in software engineering}

While a majority of the research community agrees that discretization is not a safe practice as highlighted in the prior section, discretization of the dependent variable still continues to be an accepted practice in software engineering and other fields. For instance, many software engineering studies like~\cite{guo2010characterizing,gay2010automatically,jiang2013will,schumann2009software,jalali2008optimizing,wang2017,Tian:2015,hassan2018studying} discretize the dependent variable to generate the outcome classes on which machine learning classifiers are trained.  Of these, some of the studies like~\cite{wang2017,Tian:2015,hassan2018studying} discard the data around the discretization threshold due to its ambiguous class loyalties and use only the top and bottom x\% to train classifiers. But even while doing so, they do not provide any clear theoretical or empirical reason for doing so. Whereas some other studies like~\cite{guo2010characterizing,gay2010automatically,jiang2013will,schumann2009software,jalali2008optimizing} split the continuous dependent variable on the various criterion and include all the data points without any consideration for the noise around the discretization threshold.

Therefore, our study is the first study in the field of software engineering to investigate the effect and impact of discretization noise on the performance and interpretation of classifiers. For the researchers and practitioners who continue to discretize the dependent variable artificially to build classifiers, we propose a framework. They can use our proposed framework to analyse the impact that the generated discretization noise has on their classifiers. Furthermore, we also shine the spotlight on the noise that is generated due to the discretization of the dependent variable for building classifiers - which generally is ignored in the field of software engineering.

\section{Framework for Understanding the Impact of Discretization Noise}~\label{sec:framework}
\begin{table*}[htbp]
  \centering

\caption{Estimated discretization threshold, limits and \% of data points in the noisy area for the datasets considered in the study. }
\scriptsize
   \begin{threeparttable}
    \begin{tabular}{p{1cm}|p{1cm}|p{1cm}|p{1cm}|p{1cm}|p{1cm}|p{1cm}|p{1cm}|p{1cm}|p{1cm}|p{1cm}}
    \hline
    \multirow{3}[2]{*}{\textbf{Dataset}} & \multicolumn{3}{c|}{\textbf{MT}} & \multicolumn{3}{c|}{\textbf{CT}} & \multicolumn{3}{c|}{\textbf{RTT}} & \multirow{3}[2]{*}{\textbf{\textit{step\_size}}}\\
    
   \cline{2-10}
    & \textbf{Threshold}& \textbf{Noisy area (\%)} & \textbf{Limit} &  \textbf{Threshold} & \textbf{Noisy area (\%)} & \textbf{Limit} &  \textbf{Threshold} &\textbf{Noisy area (\%)} & \textbf{Limit} & \\
    \hline
    \textbf{SO} & \multicolumn{1}{r|}{ 21.83 Mins } & \multicolumn{1}{r|}{ 29 } & \multicolumn{1}{r|}{ 55 } & \multicolumn{1}{r|}{ 136.18 Mins } & \multicolumn{1}{r|}{ 34 } & \multicolumn{1}{r|}{ 85 } & \multicolumn{1}{r|}{ 214.81 Mins } & \multicolumn{1}{r|}{ 53 } & \multicolumn{1}{r|}{ 95} & \multicolumn{1}{r}{5}\\
    \textbf{MA} & \multicolumn{1}{r|}{ 30.28 Mins } & \multicolumn{1}{r|}{ 41 } & \multicolumn{1}{r|}{ 70 } & \multicolumn{1}{r|}{ 154.48 Mins } & \multicolumn{1}{r|}{ 38 } & \multicolumn{1}{r|}{ 85 } & \multicolumn{1}{r|}{ 502.98 Mins } & \multicolumn{1}{r|}{ 44 } & \multicolumn{1}{r|}{ 95} & \multicolumn{1}{r}{5}\\
    \textbf{AU} & \multicolumn{1}{r|}{ 39.74 Mins } & \multicolumn{1}{r|}{ 38 } & \multicolumn{1}{r|}{ 70 } & \multicolumn{1}{r|}{ 329.28 Mins } & \multicolumn{1}{r|}{ 54 } & \multicolumn{1}{r|}{ 95 } & \multicolumn{1}{r|}{ 338.93 Mins } & \multicolumn{1}{r|}{ 53 } & \multicolumn{1}{r|}{ 95} & \multicolumn{1}{r}{5}\\
    \textbf{SU}  & \multicolumn{1}{r|}{ 30.14 Mins } & \multicolumn{1}{r|}{ 31 } & \multicolumn{1}{r|}{ 60 } & \multicolumn{1}{r|}{ 193.06 Mins } & \multicolumn{1}{r|}{ 62 } & \multicolumn{1}{r|}{ 95 } & \multicolumn{1}{r|}{ 262.53 Mins } & \multicolumn{1}{r|}{ 56 } & \multicolumn{1}{r|}{ 95}& \multicolumn{1}{r}{5} \\
    \textbf{PH}  & \multicolumn{1}{r|}{ 1.31 Days } & \multicolumn{1}{r|}{ 10 } & \multicolumn{1}{r|}{ 30 } & \multicolumn{1}{r|}{ 0.06 Days } & \multicolumn{1}{r|}{ 4  } & \multicolumn{1}{r|}{ 40 } & \multicolumn{1}{r|}{ 9.48 Days } & \multicolumn{1}{r|}{ 22 } & \multicolumn{1}{r|}{ 60} & \multicolumn{1}{r}{5}\\
    \textbf{BD}  & \multicolumn{1}{r|}{ 0.67 Days } & \multicolumn{1}{r|}{ 6  } & \multicolumn{1}{r|}{ 50 } & \multicolumn{1}{r|}{ 0 Days } & \multicolumn{1}{r|}{ * } & \multicolumn{1}{r|}{ * } & \multicolumn{1}{r|}{ 47.29 Days } & \multicolumn{1}{r|}{ 54 } & \multicolumn{1}{r|}{ 100} & \multicolumn{1}{r}{5}\\
    \textbf{AR}  & \multicolumn{1}{r|}{ 4.03 Stars } & \multicolumn{1}{r|}{ 43 } & \multicolumn{1}{r|}{ 7  } & \multicolumn{1}{r|}{ 3.86 Stars } & \multicolumn{1}{r|}{ 50 } & \multicolumn{1}{r|}{ 10 } & \multicolumn{1}{r|}{ 3.74 Stars } & \multicolumn{1}{r|}{ 63 } & \multicolumn{1}{r|}{ 15} & \multicolumn{1}{r}{0.5}\\
    \hline
    \end{tabular}%
        \begin{tablenotes}
      \scriptsize
      \item *The automated noisy area estimation algorithm found no data points in the noisy area
      \item \textbf{MT}- \textbf{M}edian based discretization \textbf{T}hreshold, \textbf{CT}- Univariate \textbf{C}lustering based discretization \textbf{T}hreshold, \textbf{RTT}- CA\textbf{RT} based discretization \textbf{t}hreshold
      \item \textbf{Datasets:} SO- Stack Overflow, MA- Mathematics, AU- Ask Ubuntu, SU- Super User, PH- Patch, BD- Bug-delay, AR- App-rating
    \end{tablenotes}

\end{threeparttable}
  \label{tab:limits}%
\end{table*}%

\begin{figure*}
    \includegraphics[width=\linewidth,scale=1.5]{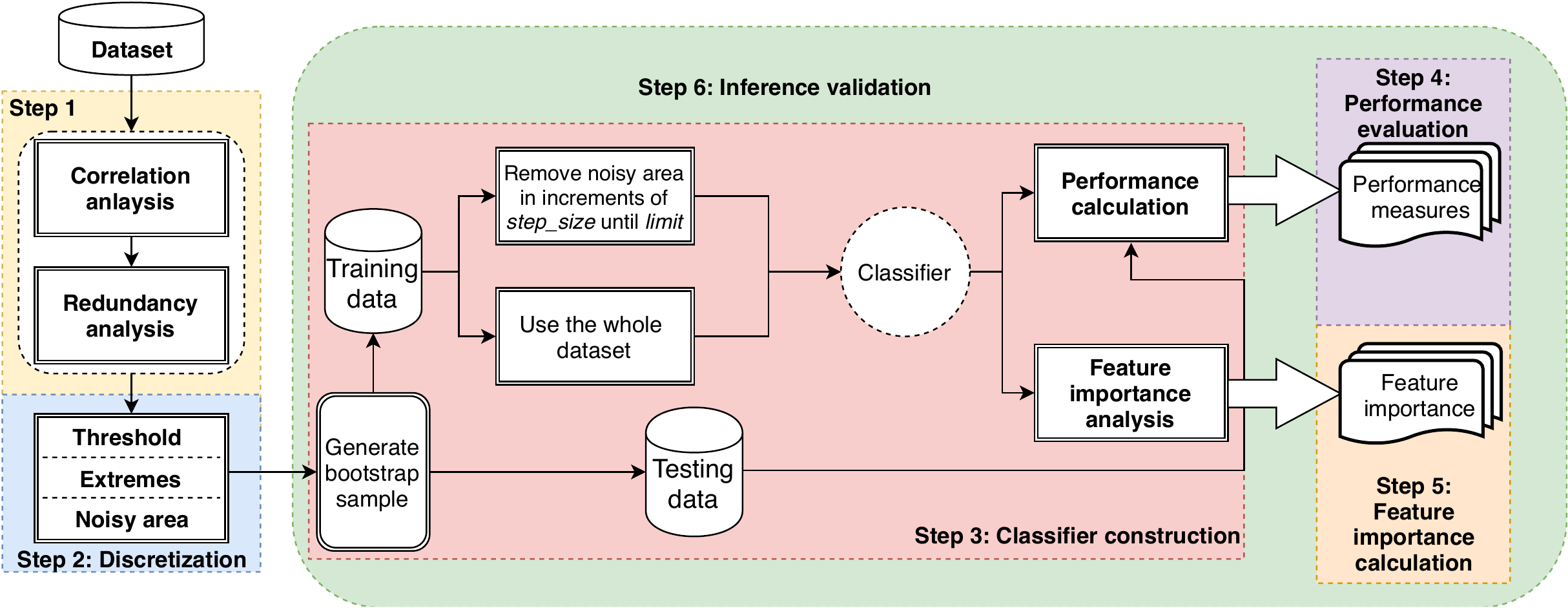}

    \caption{Overview of our framework.}
    \label{fig:fig1}
\end{figure*}

An overview of our framework for understanding the impact of discretization noise is presented in Figure~\ref{fig:fig1}. The framework consists of six steps.
Due to space constraints, we detail the steps below (with more explanations where needed) with a running example in Appendix~B to better demonstrate the use of our framework.

The individual steps of our approach are explained in detail below.

\subsection{Step 1: Correlation \& Redundancy Analysis}

In this study, we collect data based on two criteria: 1) the dependent variable is continuous; 2) the dependent variable lacks a clear-cut threshold for discretization. Based on these two criteria, we collect 4 types of data (7 datasets): Q\&A websites data~\cite{wang2017}, Linux patch acceptance time (Patch) data~\cite{jiang2013will}, Bug-fix delay time (Bug-delay) data~\cite{zhang2012empirical}, and Mobile app rating (App-rating) data~\cite{Tian:2015} (More details about the datasets are given in Appendix~A). Table~\ref{tab:dataset_details} contains basic information about the number of data points and the independent features of each of the studied datasets.

We perform correlation and redundancy analysis on the independent features of a studied dataset to remove correlated and redundant features from the dataset, thereby not biasing our feature importance results~\cite{tantithamthavorn2018experience}. We do so instead of using other common and state of the art dimensionality reduction techniques like PCA, since, dimensionality reduction techniques like PCA combine and transform the original features into principal components, which are no longer directly interpretable. Finally, a recent MSR study by Ghothra~\etal~\cite{ghotra2017large} showed that correlation-based feature selection is very robust for software engineering datasets.  Though we use and recommend correlation and redundancy analysis, our framework supports the use of other methods. We do not pre-process the independent features of the dataset any further in our study. However, practitioners can perform other data pre-processing steps like imputation if required.

\subsection{Step 2: Discretization}\label{sec:discretization}

\textbf{Threshold estimation:} The primary objective of our framework is to understand the impact of discretization noise. We discretize the dependent variable with respect to an artificial threshold (a.k.a a cutpoint) into two response classes: ``class1'' and ``class2''. We then assign the ``class1'' class label to all the data points with a dependent variable that has a value that is less than or equal to the chosen discretization threshold (e.g., median). The remaining data points are assigned the class label ``class2''.
The artificial threshold for such a discretization could be chosen in multiple ways. The threshold could be domain specific and be defined by the experts (e.g., ideal bug-fix time for a specific project as defined by the software engineers working on the project). Alternatively, in the absence of such an established domain specific discretization threshold, many of the prior studies have resorted to various heuristic, intuitive and alternate thresholds for discretization~\cite{altman2006cost,de1998investigation,el2001modelling,Tian:2015,wang2017}. But irrespective of the choice of the discretization threshold, the data points close to the discretization threshold produce discretization noise. Our framework analyses the impact of discretization noise generated by any such discretization threshold.

In this study, to demonstrate the generalizability and applicability of our framework, we use three artificial discretization thresholds. 

\noindent\textit{\textbf{M}edian based discretization \textbf{T}hreshold (\textbf{MT}):~}Many prior studies use median for discretizing the dependent variable into binary classes~\cite{de1998investigation,el2001modelling} and it is often used in the absence of explicit domain knowledge about the classes of a dependent variable~\cite{altman2006cost}.

\noindent\textit{Univariate \textbf{C}lustering based \textbf{T}hreshold (\textbf{CT}):~}Univariate clustering is an automated technique for discretization. Univariate clustering splits the dependent variable into multiple groups in an optimal fashion. We use Wang and Song's implementation \textit{optimal k-means clustering in one dimension} (ckmeans.1d.dp\footnote{https://cran.r-project.org/web/packages/Ckmeans.1d.dp/index.html}) here~\cite{wang2011ckmeans}. The ckmeans.1d.dp divides data in one dimension into k clusters so that the sum of squares of within-cluster distances from each element to its corresponding cluster mean is minimized~\cite{wang2011ckmeans}.  We set k equal 2 since we wish to divide the dependent variable into two classes.

\noindent\textit{CA\textbf{RT} based discretization \textbf{T}hreshold (\textbf{RTT}):~} We use the regression tree approach as described by Breiman~\cite{breiman2017classification}.\footnote{https://cran.r-project.org/web/packages/rpart/index.html} Here, we use the continuous dependent variable of our dataset as both the independent and the target variable for the regression tree. We then use the generated regression tree's root node as the threshold for discretization since we attempt to split the dependent variable into two classes.

The generated discretization threshold is used for discretizing the continuous dependent variable into binary classes.

\noindent\textbf{Noisy area estimation:} Once the dataset is discretized, we need to define the area of the dataset which contains discretization noise as the noisy area. Domain experts could determine a specific range of values around the discretization threshold to be noisy and this could be used as the noisy area. But as we lack deep domain expertise of the datasets considered in this study (which might be the case for many practitioners), we present Algorithm~\ref{algo1}: an automated algorithm for estimating the noisy area in a given dataset.

We define the \textit{noisy area} as the data points whose class loyalties are hard to discern due to their proximity to the artificial discretization threshold. Such a hypothesis follows from the rationale of prior studies where the data points around the discretization threshold were discarded to provide better class separation in the training data~\cite{Tian:2015,wang2017,judd2009learning,abdelwahab2015supervised,abdelmoez2012bug}. Algorithm~\ref{algo1} takes the dataset, cutpoint (i.e., the discretization threshold), and \textit{step\_size} as input parameters. \textit{step\textunderscore size} controls the granularity of the analysis (i.e., the size of the increment from the cut point) - a smaller value of the \textit{step\textunderscore size} allows for a finer estimation of the noisy area, whereas a larger value provides a coarse estimation of the noisy area. The \textit{step\_size} used for all the datasets in this study is given the Table~\ref{tab:limits}.

Line 1 to 3 of the algorithm establishes the initial candidate noisy area, by selecting the area around the cutpoint. More specifically, we consider the points within the area $cutpoint \pm cutpoint *100\%$ as the candidate noisy area. We do so for two reasons: 1) most of the discretization noise would be concentrated around the discretization threshold due to its proximity to the threshold. 2) if we consider more data, we might not be able to ascertain if the impact of the noisy area on the performance and interpretation of a classifier is due to discretization noise or the high volume of data that is lost. Through line 4 to line 9 we incrementally subset the dataset into the quantum of size given by $cutpoint \pm cutpoint *setp\textunderscore size$ and compute the non-linearity of the quantum.

\begin{figure}
    \includegraphics[width=\linewidth]{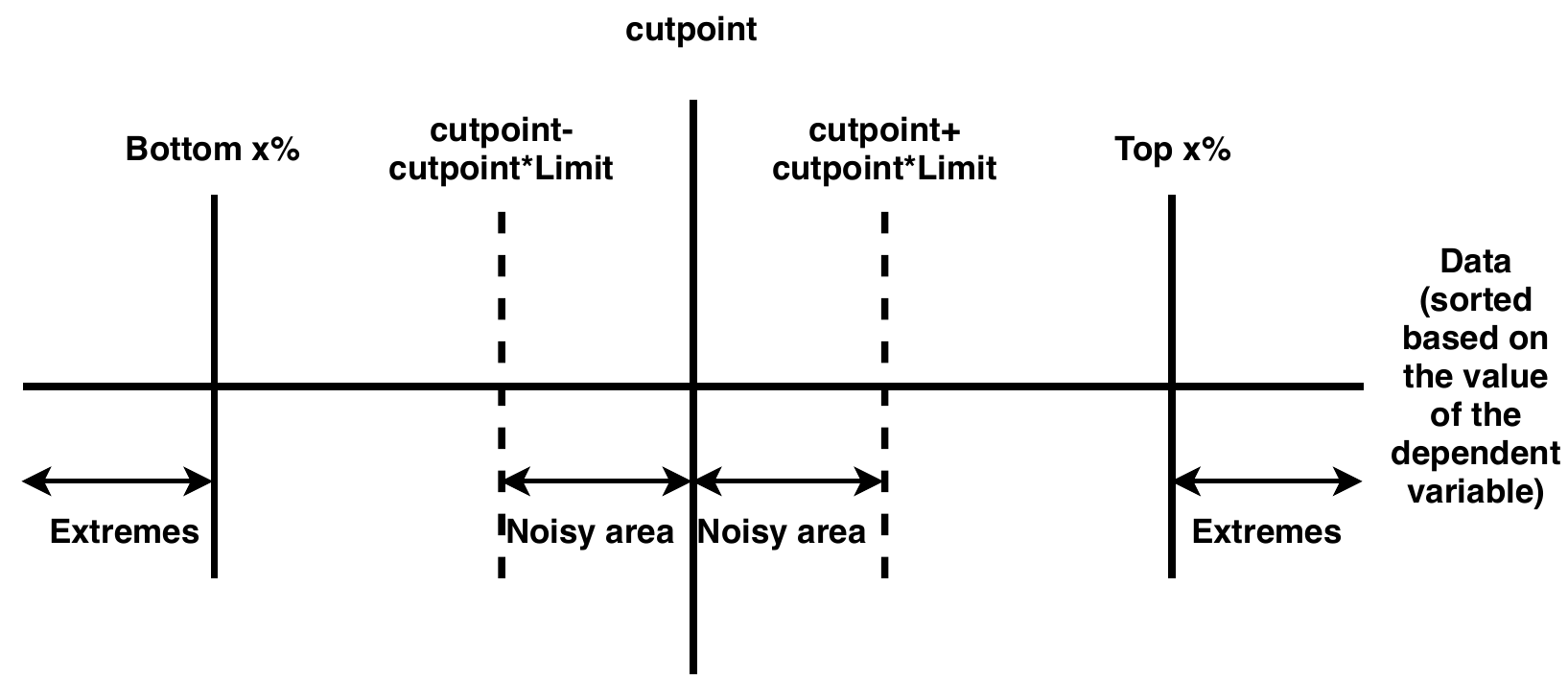}
    \caption{Extremes and noisy area definitions of a dataset.}
    \label{fig:framework_2}
\end{figure}

Non-linearity is one of the complexity measures defined by Ho and Basu~\cite{ho2002complexity}. Non-linearity score attempts to quantify how hard it might be for a classifier to classify the data points (please refer Section~\ref{sec:disc} and Table 4 in the appendix for more details about complexity measures). Once we establish the non-linearity for all the quanta, we take the quantum with the maximum non-linearity as the noisy area for our analysis and the step\textunderscore size that yielded the quantum as the $limit$, which we use to demarcate the noisy area. We use the maximum non-linearity to demarcate the noisy area as it indicates the quantum with the highest data complexity (thereby harder for the classifier). Figure~\ref{fig:framework_2} demonstrates how the limit value is used to demarcate the noisy area in a dataset. We present the $limit$ that is generated for demarcating the noisy area of all the studied datasets for various discretization thresholds in Table~\ref{tab:limits}.

\noindent\textbf{Extremes estimation:} Finally we establish the data points with the least discretization noise as the extremes. These data points typically have high discriminative power as they are the furthest away from the discretization threshold. Extremes are typically the data points that are associated with the top and bottom x\% of the sorted continuous dependent variable. Prior studies usually consider the top and bottom $x\%$ as data points that are devoid of noise and use them for constructing the classifier~\cite{wang2017,Tian:2015}. We use $x$ as 10\% in this study. But the framework allows using any value without any further change to the overall methodology.

\begin{algorithm}

\scriptsize
\SetAlgoLined
\KwIn{\textit{dataset, cutpoint, step\textunderscore size}}
\KwOut{\textit{limit}}
\KwResult{Estimates the noisy area in the data automatically by computing the limit}
 $lower\textunderscore limit = cutpoint - cutpoint*100\%$\\
 $upper\textunderscore limit = cutpoint + cutpoint*100\%$\\

$ noisy\textunderscore area = SUBSET(dataset,lower\textunderscore limit$,$upper\textunderscore limit)$\\

 \While{$((cutpoint \pm cutpoint*step\textunderscore size) \le upper\textunderscore limit~AND~\ge lower\textunderscore limit)$}{

$quanta = $~SUBSET$(noisy\textunderscore area,~cutpoint - cutpoint*step\textunderscore size$,~$cutpoint + cutpoint*step\textunderscore size)$\\

$nl\textunderscore score = $~COMPUTE\textunderscore NON\textunderscore LINEARITY$(quanta)$\\

$results[step\textunderscore size] = (nl\textunderscore score)$\\
$step\textunderscore size\mathrel{+}=step\textunderscore size$
 }
$limit = Index of $~MAX$(results)$
 \caption{Automated noisy area estimation algorithm}
 \label{algo1}
\end{algorithm}

\subsection{Step 3: Classifier construction}\label{sec:construction}

To study the impact of discretization noise, we construct a classifier on the whole dataset and on the dataset with the noisy area removed. One can choose any classifier of their choice in this step. In our study, we consider the 6 classifiers considered by Rajbahadur~\etal~\cite{rajbahadur2017impact}. From the 6 classifiers, we choose the classifiers that have a default feature importance computation method as our framework studies the impact of discretization noise on both the performance and feature importance. Therefore we demonstrate the capability of our framework to analyse the impact of discretization noise on random forest classifier (RFCM), Logistic Regression (LR), Classification and Regression Tree (CART), and K-Nearest Neighbour (KNN). All of the chosen classifiers are hyper parameter tuned to ensure the best and stable performance. We used the method used by  Tantithamthavorn~\etal~\cite{tantithamthavorn2016automated} to hyper-parameter tune all of our classifiers.

Though we use the four aforementioned classifiers, one can use other classifiers instead of these classifiers without any changes to the other steps in the framework.


\subsection{Step 4: Performance Evaluation}\label{measures}

In this step, the desired classifier performance evaluation measures are chosen. In this study, we observe and evaluate the performance of the constructed classifiers on \textit{Accuracy, Precision, Recall, Brier score, Area Under the receiver operator characteristic Curve (AUC), F-measure, and Mathew's Correlation Coefficient (MCC)}, since many prior studies studied the performance of classifiers using these measures~\cite{zhang2014towards,boughorbel2017optimal,brier1950verification}. We calculate these measures with ``class 1'' as the relevant (positive) class. 


Though we demonstrate our framework on the aforementioned performance measures, our framework allows users to use any performance evaluation measures (by themselves or in combination with other measures).


\subsection{Step 5: Feature Importance Calculation}\label{sec:feature_impo_estimation}
We use the default feature importance calculation technique that is associated with each of the studied classifiers to compute the feature importance for each classifier. We use the variable importance computation method \textbf{VarImp()} of \textbf{caret} package to compute the feature importance of the studied classifiers.


\subsection{Step 6: Inference Validation} \label{sec:bootstrap}

To ensure that the conclusions that we draw about our classifiers are statistically robust, we use the 100 out-of-sample bootstrap validation technique, which yields an optimal balance between the bias and variance as suggested in the recent study of Tantithamthavorn \textit{\etal}~\cite{tantithamthavorn2017empirical}.



The out-of-sample bootstrap process is repeated 100 times. After the bootstrap validation, 100 performance measures and 100 lists of derived feature importance ranks are generated. We carry out further analysis on these generated performance measures and the derived feature importance ranks to investigate our research questions.

\subsection{Framework Deployment}\label{sec:deployment}
We use our framework of 6 steps on any given dataset to analyse the impact of discretization noise (as demonstrated in Section~\ref{sec:rqs}) along with performance and interpretation on the chosen classifier. Step 1 removes the correlation and redundancy among the features in a dataset, while step 2 is pivotal for estimating the noisy area and extremes for a chosen discretization threshold. Steps 3 to 6 are repeated by incrementally discarding data points in increments of the $step\_size$ parameter (smaller \textit{step\_size} enables finer analysis and vice versa) in the noisy area around the threshold until all of the data points in the noisy area are discarded. Such an incremental analysis helps the framework identify the impact of discretization noise and determine the exact amount of data from the noisy area that needs to be discarded for a given dataset, discretization threshold and classifier of choice. We also provide an R package\footnote{https://github.com/SAILResearch/suppmaterial-19-gopi-discretization\_noise\_impact}\footnote{ Username: discretizationreviewer Password: DiscNoise123} of our framework to enable others and practitioners automated support to use our framework with trivial effort.

\section{Understanding the Impact of Discretization Noise on the Performance and Interpretation of a Classifier}\label{sec:rqs}

\subsection{\textbf{Studying the Impact of Discretization Noise on the Performance of a Classifier}}\label{sec:rq1}




\noindent\textbf{Motivation:}
It is intuitive to expect that discretization noise might impact the performance of a classifier. Ferri~\etal~ show that different performance measures are impacted differently by different types of noise in a dataset~\cite{ferri2009experimental}.
Therefore, first, it is essential to establish if discretization noise impacts the performance of a classifier like other noises. Second, if the discretization noise does impact the performance of a classifier, we need to analyse how the discretization noise in a dataset impacts the performance of a classifier (either positively/negatively) in different performance measures. Finally, it is essential to establish how much data do we have to discard to avoid the impact of discretization noise (as opposed to using only the top and bottom x\%).

In order to enable researchers and practitioners to perform such an analysis in a generalizable fashion, we propose our framework. Our framework enables researchers and practitioners to examine the impact of discretization noise on the performance of various classifiers using a variety of software engineering datasets, across a multitude of performance measures. \\

\begin{table*}[htbp]
  \centering
\scriptsize
  \caption{Percentage of improvement in median performance of various classifiers with the noisy area removed over classifiers with no data removed across various performance measures (The $x$ value for which the performance impact first occurs for the given measure is also provided).}
  \begin{threeparttable}
    \begin{tabular}{c|p{1cm}|p{0.05cm}|p{0.05cm}|p{0.05cm}|p{0.05cm}|p{0.05cm}|p{0.05cm}|p{0.05cm}|p{0.05cm}|p{0.05cm}|p{0.05cm}|p{0.05cm}|p{0.05cm}|p{0.05cm}|p{0.05cm}|}
    \hline
    \multirow{2}[2]{*}{\textbf{Classifier}} & \multirow{2}[2]{*}{\textbf{Dataset}} & \multicolumn{2}{c|}{\textbf{ACC (\%)}} & \multicolumn{2}{c|}{\textbf{PRC (\%)}} & \multicolumn{2}{c|}{\textbf{RCL (\%)}} & \multicolumn{2}{c|}{\textbf{BS (\%)}} & \multicolumn{2}{c|}{\textbf{AUC (\%)}} & \multicolumn{2}{c|}{\textbf{F-M (\%)}} & \multicolumn{2}{c}{\textbf{MCC (\%)}} \\
    \cline{3-16}
\multirow{8}[2]{*}{RF} &  & \multicolumn{1}{c|}{ \textbf{Mag}} & \multicolumn{1}{c|}{$x$} & \multicolumn{1}{c|}{ \textbf{Mag}} & \multicolumn{1}{c|}{$x$}& \multicolumn{1}{c|}{ \textbf{Mag}} & \multicolumn{1}{c|}{$x$}& \multicolumn{1}{c|}{ \textbf{Mag}} & \multicolumn{1}{c|}{$x$}& \multicolumn{1}{c|}{ \textbf{Mag}} & \multicolumn{1}{c|}{$x$}& \multicolumn{1}{c|}{ \textbf{Mag}} & \multicolumn{1}{c|}{$x$}& \multicolumn{1}{c|}{ \textbf{Mag}} & \multicolumn{1}{c}{$x$} \\
    \hline

        & \multicolumn{1}{r|}{ SO } & \multicolumn{1}{r|}{ \textbf{-0.65}} & \multicolumn{1}{r|}{ 50}  & \multicolumn{1}{r|}{ \textbf{5.07}} & \multicolumn{1}{r|}{ 15}  & \multicolumn{1}{r|}{ \textbf{-12.42}} & \multicolumn{1}{r|}{ 15} & \multicolumn{1}{r|}{ \textbf{-8}} & \multicolumn{1}{r|}{ 5}  & \multicolumn{1}{r|}{ 0} & \multicolumn{1}{r|}{ 0 } & \multicolumn{1}{r|}{ \textbf{-3.69}} & \multicolumn{1}{r|}{ 25}  & \multicolumn{1}{r|}{ -0.054\textsuperscript{\#}} & \multicolumn{1}{r}{ 55}  \\
        & \multicolumn{1}{r|}{ MA } & \multicolumn{1}{r|}{ \textbf{-3.5}} & \multicolumn{1}{r|}{ 45}  & \multicolumn{1}{r|}{ \textbf{8.96}} & \multicolumn{1}{r|}{ 15}  & \multicolumn{1}{r|}{ \textbf{-32.2}} & \multicolumn{1}{r|}{ 15} & \multicolumn{1}{r|}{ \textbf{-11.77}} & \multicolumn{1}{r|}{ 5}  & \multicolumn{1}{r|}{ -1.23\textsuperscript{\#}} & \multicolumn{1}{r|}{ 70 } & \multicolumn{1}{r|}{ \textbf{-11.61}} & \multicolumn{1}{r|}{ 20}  & \multicolumn{1}{r|}{ \textbf{-6.38}} & \multicolumn{1}{r}{ 50 } \\
        & \multicolumn{1}{r|}{ AU } & \multicolumn{1}{r|}{ \textbf{-7.76}} & \multicolumn{1}{r|}{ 40} & \multicolumn{1}{r|}{ \textbf{12.09}} & \multicolumn{1}{r|}{ 30} & \multicolumn{1}{r|}{ \textbf{-99.02}} & \multicolumn{1}{r|}{ 25} & \multicolumn{1}{r|}{ \textbf{-7.79}} & \multicolumn{1}{r|}{ 5}  & \multicolumn{1}{r|}{ 0} & \multicolumn{1}{r|}{ 0 } & \multicolumn{1}{r|}{ \textbf{-43.19}} & \multicolumn{1}{r|}{ 25}  & \multicolumn{1}{r|}{ \textbf{-18.64}} & \multicolumn{1}{r}{ 45} \\
        & \multicolumn{1}{r|}{ SU } & \multicolumn{1}{r|}{ -0.53\textsuperscript{\S}} & \multicolumn{1}{r|}{ 0 } & \multicolumn{1}{r|}{ \textbf{5.88}} & \multicolumn{1}{r|}{ 25} & \multicolumn{1}{r|}{ \textbf{-20.4}} & \multicolumn{1}{r|}{ 20} & \multicolumn{1}{r|}{ \textbf{-7.82}} & \multicolumn{1}{r|}{ 5}  & \multicolumn{1}{r|}{ 0\textsuperscript{\S}} & \multicolumn{1}{r|}{ 45 } & \multicolumn{1}{r|}{ \textbf{-7.59}} & \multicolumn{1}{r|}{ 30}  & \multicolumn{1}{r|}{ 0.36} & \multicolumn{1}{r}{ 0 }      \\
        & \multicolumn{1}{r|}{ PH } & \multicolumn{1}{r|}{ \textbf{-1.64}} & \multicolumn{1}{r|}{ 10} & \multicolumn{1}{r|}{ \textbf{1.36}} & \multicolumn{1}{r|}{ 25} & \multicolumn{1}{r|}{ \textbf{-6.64}} & \multicolumn{1}{r|}{ 10} & \multicolumn{1}{r|}{ \textbf{-2.04}} & \multicolumn{1}{r|}{ 5} & \multicolumn{1}{r|}{ \textbf{-2.22}} & \multicolumn{1}{r|}{ 10} & \multicolumn{1}{r|}{ \textbf{-2.62}} & \multicolumn{1}{r|}{ 10}  & \multicolumn{1}{r|}{ \textbf{-4.08}} & \multicolumn{1}{r}{ 10 } \\
        & \multicolumn{1}{r|}{ BD } & \multicolumn{1}{r|}{ 0.47\textsuperscript{\S}} & \multicolumn{1}{r|}{ 0 } & \multicolumn{1}{r|}{ 0.72} & \multicolumn{1}{r|}{ 0 } & \multicolumn{1}{r|}{ 0.75} & \multicolumn{1}{r|}{ 0 } & \multicolumn{1}{r|}{ -0.66\textsuperscript{\#}} & \multicolumn{1}{r|}{ 40 } & \multicolumn{1}{r|}{ 0\textsuperscript{\S}} & \multicolumn{1}{r|}{ 0 } & \multicolumn{1}{r|}{ 0.74\textsuperscript{\S}} & \multicolumn{1}{r|}{ 0 } & \multicolumn{1}{r|}{ 1.81\textsuperscript{\S}} & \multicolumn{1}{r}{ 0 }   \\
        & \multicolumn{1}{r|}{ AR } & \multicolumn{1}{r|}{ -0.78\textsuperscript{\S}} & \multicolumn{1}{r|}{ 0 } & \multicolumn{1}{r|}{ \textbf{-4.23}} & \multicolumn{1}{r|}{ 3}& \multicolumn{1}{r|}{ \textbf{13.17}} & \multicolumn{1}{r|}{ 2} & \multicolumn{1}{r|}{ \textbf{-2.89}} & \multicolumn{1}{r|}{ 0.5} & \multicolumn{1}{r|}{ 0} & \multicolumn{1}{r|}{ 0 } & \multicolumn{1}{r|}{ \textbf{4.52}} & \multicolumn{1}{r|}{ 2} & \multicolumn{1}{r|}{ -2.93\textsuperscript{\S}} & \multicolumn{1}{r}{ 0 } \\
      \hline
          \multirow{7}[2]{*}{LR} & \multicolumn{1}{r|}{SO} & \multicolumn{1}{r|}{ -0.76\textsuperscript{\#}} & \multicolumn{1}{r|}{ 55 } & \multicolumn{1}{r|}{ \textbf{6.24}} & \multicolumn{1}{r|}{ 15}  & \multicolumn{1}{r|}{ \textbf{-20.81}} & \multicolumn{1}{r|}{ 20} & \multicolumn{1}{r|}{ \textbf{-7.73}} & \multicolumn{1}{r|}{ 5}  & \multicolumn{1}{r|}{ 0\textsuperscript{\S}} & \multicolumn{1}{r|}{ 0 } & \multicolumn{1}{r|}{ \textbf{-7.04}} & \multicolumn{1}{r|}{ 25}  & \multicolumn{1}{r|}{  -0.83} & \multicolumn{1}{r}{ 0 }    \\
        & \multicolumn{1}{r|}{ MA } & \multicolumn{1}{r|}{ \textbf{-1.78}} & \multicolumn{1}{r|}{ 15} & \multicolumn{1}{r|}{ \textbf{12.33}} & \multicolumn{1}{r|}{ 10} & \multicolumn{1}{r|}{ \textbf{-46.87}} & \multicolumn{1}{r|}{ 10}  & \multicolumn{1}{r|}{ \textbf{-12.02}} & \multicolumn{1}{r|}{ 5} & \multicolumn{1}{r|}{ 0\textsuperscript{\#}} & \multicolumn{1}{r|}{ 70 } & \multicolumn{1}{r|}{ \textbf{-15.28}} & \multicolumn{1}{r|}{ 15}  & \multicolumn{1}{r|}{ \textbf{ -2.59}} & \multicolumn{1}{r}{ 25}\\
        & \multicolumn{1}{r|}{ AU } & \multicolumn{1}{r|}{ \textbf{-5.54}} & \multicolumn{1}{r|}{ 50} & \multicolumn{1}{r|}{ \textbf{11.47}} & \multicolumn{1}{r|}{ 30} & \multicolumn{1}{r|}{ \textbf{-139.29}} & \multicolumn{1}{r|}{ 25} & \multicolumn{1}{r|}{ \textbf{-10.42}} & \multicolumn{1}{r|}{ 10} & \multicolumn{1}{r|}{ 1.56\textsuperscript{\S}} & \multicolumn{1}{r|}{ 0 } & \multicolumn{1}{r|}{ \textbf{-59.78}} & \multicolumn{1}{r|}{ 25}  & \multicolumn{1}{r|}{ \textbf{ -16.3}} & \multicolumn{1}{r}{ 60} \\
        & \multicolumn{1}{r|}{ SU } & \multicolumn{1}{r|}{ \textbf{-1.17}} & \multicolumn{1}{r|}{ 60} & \multicolumn{1}{r|}{ \textbf{6.42}} & \multicolumn{1}{r|}{ 30}  & \multicolumn{1}{r|}{ \textbf{-48.2}} & \multicolumn{1}{r|}{ 20}  & \multicolumn{1}{r|}{ \textbf{-5.87}} & \multicolumn{1}{r|}{ 10} & \multicolumn{1}{r|}{ 0} & \multicolumn{1}{r|}{ 0 } & \multicolumn{1}{r|}{ \textbf{-18.92}} & \multicolumn{1}{r|}{ 20}  & \multicolumn{1}{r|}{ \textbf{ -4.61}} & \multicolumn{1}{r}{ 60} \\
        & \multicolumn{1}{r|}{ PH } & \multicolumn{1}{r|}{ -0.08\textsuperscript{\S}} & \multicolumn{1}{r|}{ 0 } & \multicolumn{1}{r|}{ \textbf{2.23}} & \multicolumn{1}{r|}{ 15} & \multicolumn{1}{r|}{ \textbf{-4.68}} & \multicolumn{1}{r|}{ 10} & \multicolumn{1}{r|}{ \textbf{-2.59}} & \multicolumn{1}{r|}{ 5} & \multicolumn{1}{r|}{ 0    \textsuperscript{\S}} & \multicolumn{1}{r|}{ 0 } & \multicolumn{1}{r|}{ \textbf{-0.97}} & \multicolumn{1}{r|}{ 15} & \multicolumn{1}{r|}{ \textbf{ -1.28}} & \multicolumn{1}{r}{ 20} \\
        & \multicolumn{1}{r|}{ BD } & \multicolumn{1}{r|}{ -0.2} & \multicolumn{1}{r|}{ 0 } & \multicolumn{1}{r|}{ -0.39} & \multicolumn{1}{r|}{ 0 } & \multicolumn{1}{r|}{ 1.63\textsuperscript{\S}} & \multicolumn{1}{r|}{ 0 } & \multicolumn{1}{r|}{-0.78\textsuperscript{\#}} & \multicolumn{1}{r|}{ 40 } & \multicolumn{1}{r|}{ 0} & \multicolumn{1}{r|}{ 0 } & \multicolumn{1}{r|}{ 0.98} & \multicolumn{1}{r|}{ 0 } & \multicolumn{1}{r|}{  -0.18} & \multicolumn{1}{r}{ 0}     \\
        & \multicolumn{1}{r|}{ AR } & \multicolumn{1}{r|}{ -0.36\textsuperscript{\S}} & \multicolumn{1}{r|}{ 0 } & \multicolumn{1}{r|}{ \textbf{-2.83}} & \multicolumn{1}{r|}{ 4} & \multicolumn{1}{r|}{ \textbf{10.82}} & \multicolumn{1}{r|}{ 2} & \multicolumn{1}{r|}{ \textbf{-8.03}} & \multicolumn{1}{r|}{ 0.5} & \multicolumn{1}{r|}{ 0} & \multicolumn{1}{r|}{ 0 } & \multicolumn{1}{r|}{ \textbf{3.95}} & \multicolumn{1}{r|}{ 2} & \multicolumn{1}{r|}{  0.44} & \multicolumn{1}{r}{ 0  }   \\
    \hline
      \multirow{7}[2]{*}{CART}  & \multicolumn{1}{r|}{SO} & \multicolumn{1}{r|}{ \textbf{1.41}} & \multicolumn{1}{r|}{ 10}  & \multicolumn{1}{r|}{ \textbf{8.58}} & \multicolumn{1}{r|}{ 15} & \multicolumn{1}{r|}{ \textbf{-16.77}} & \multicolumn{1}{r|}{ 20}& \multicolumn{1}{r|}{ -2.77\textsuperscript{\#}} & \multicolumn{1}{r|}{ 30 } & \multicolumn{1}{r|}{ \textbf{3.85}} & \multicolumn{1}{r|}{ 20} & \multicolumn{1}{r|}{ \textbf{-3.42}} & \multicolumn{1}{r|}{ 35} & \multicolumn{1}{r|}{ \textbf{5.36}} & \multicolumn{1}{r}{ 15 } \\
        & \multicolumn{1}{r|}{ MA } & \multicolumn{1}{r|}{ \textbf{-1.67}} & \multicolumn{1}{r|}{ 10} & \multicolumn{1}{r|}{ \textbf{11.26}} & \multicolumn{1}{r|}{ 20}  & \multicolumn{1}{r|}{ \textbf{-37.16}} & \multicolumn{1}{r|}{ 20} & \multicolumn{1}{r|}{ \textbf{-6.54}} & \multicolumn{1}{r|}{ 60}  & \multicolumn{1}{r|}{ 0} & \multicolumn{1}{r|}{ 30 } & \multicolumn{1}{r|}{ \textbf{-11.93}} & \multicolumn{1}{r|}{ 30}  & \multicolumn{1}{r|}{ -1.27\textsuperscript{\#}} & \multicolumn{1}{r}{ 10} \\
        & \multicolumn{1}{r|}{ AU } & \multicolumn{1}{r|}{ -0.56\textsuperscript{\S}} & \multicolumn{1}{r|}{ 0 } & \multicolumn{1}{r|}{ \textbf{7.34}} & \multicolumn{1}{r|}{ 30} & \multicolumn{1}{r|}{ \textbf{-40.49}} & \multicolumn{1}{r|}{ 30} & \multicolumn{1}{r|}{ \textbf{-9.55}} & \multicolumn{1}{r|}{ 50} & \multicolumn{1}{r|}{ 0} & \multicolumn{1}{r|}{ 50 } & \multicolumn{1}{r|}{ \textbf{-15.97}} & \multicolumn{1}{r|}{ 40} & \multicolumn{1}{r|}{ 2.47\textsuperscript{\S}} & \multicolumn{1}{r}{ 0}   \\
        & \multicolumn{1}{r|}{ SU } & \multicolumn{1}{r|}{ \textbf{1.46}} & \multicolumn{1}{r|}{ 20} & \multicolumn{1}{r|}{ \textbf{6.93}} & \multicolumn{1}{r|}{ 25} & \multicolumn{1}{r|}{ \textbf{-21.36}} & \multicolumn{1}{r|}{ 45} & \multicolumn{1}{r|}{ \textbf{-7.32}} & \multicolumn{1}{r|}{ 55}  & \multicolumn{1}{r|}{ \textbf{2.99}} & \multicolumn{1}{r|}{ 25} & \multicolumn{1}{r|}{ \textbf{-7.05}} & \multicolumn{1}{r|}{ 50} & \multicolumn{1}{r|}{ \textbf{9.52}} & \multicolumn{1}{r}{ 20 } \\
        & \multicolumn{1}{r|}{ PH } & \multicolumn{1}{r|}{ \textbf{-1.12}} & \multicolumn{1}{r|}{ 10}  & \multicolumn{1}{r|}{ \textbf{1.76}} & \multicolumn{1}{r|}{ 25} & \multicolumn{1}{r|}{ \textbf{-6.77}} & \multicolumn{1}{r|}{ 15} & \multicolumn{1}{r|}{ -1.79\textsuperscript{\S}} & \multicolumn{1}{r|}{ 0 } & \multicolumn{1}{r|}{ \textbf{-1.79}} & \multicolumn{1}{r|}{ 15} & \multicolumn{1}{r|}{ \textbf{-2.45}} & \multicolumn{1}{r|}{ 10} & \multicolumn{1}{r|}{ \textbf{-3.21}} & \multicolumn{1}{r}{ 10}\\
        & \multicolumn{1}{r|}{ BD } & \multicolumn{1}{r|}{ 1.2\textsuperscript{\S}} & \multicolumn{1}{r|}{ 0 } & \multicolumn{1}{r|}{ 1.33\textsuperscript{\S}} & \multicolumn{1}{r|}{ 0 } & \multicolumn{1}{r|}{ 0.67} & \multicolumn{1}{r|}{ 0 } & \multicolumn{1}{r|}{ -0.77\textsuperscript{\S}} & \multicolumn{1}{r|}{ 0 } & \multicolumn{1}{r|}{ 1.54\textsuperscript{\S}} & \multicolumn{1}{r|}{ 0 } & \multicolumn{1}{r|}{ 1.69\textsuperscript{\S}} & \multicolumn{1}{r|}{ 0 } & \multicolumn{1}{r|}{ 6.53\textsuperscript{\S}} & \multicolumn{1}{r}{ 0}   \\
        & \multicolumn{1}{r|}{ AR } & \multicolumn{1}{r|}{ 0.33} & \multicolumn{1}{r|}{ 0 } & \multicolumn{1}{r|}{ -1.06\textsuperscript{\#}} & \multicolumn{1}{r|}{ 6.5 } & \multicolumn{1}{r|}{ \textbf{9.92}} & \multicolumn{1}{r|}{ 3.5} & \multicolumn{1}{r|}{ -3.62\textsuperscript{\S}} & \multicolumn{1}{r|}{ 0 } & \multicolumn{1}{r|}{ 1.64\textsuperscript{\S}} & \multicolumn{1}{r|}{ 0 } & \multicolumn{1}{r|}{ \textbf{4.35}} & \multicolumn{1}{r|}{ 3.5} & \multicolumn{1}{r|}{ 3.79\textsuperscript{\S}} & \multicolumn{1}{r}{ 0 }  \\
    \hline
    \multirow{7}[2]{*}{KNN}  & \multicolumn{1}{r|}{ SO} & \multicolumn{1}{r|}{ \textbf{2.64}} & \multicolumn{1}{r|}{ 10 } & \multicolumn{1}{r|}{ \textbf{6.02}} & \multicolumn{1}{r|}{ 10 } & \multicolumn{1}{r|}{ \textbf{-9.88}} & \multicolumn{1}{r|}{ 25 } & \multicolumn{1}{r|}{ \textbf{-3.72}} & \multicolumn{1}{r|}{ 5 } & \multicolumn{1}{r|}{ \textbf{4.29}} & \multicolumn{1}{r|}{ 5 } & \multicolumn{1}{r|}{ \textbf{-1.81}} & \multicolumn{1}{r|}{ 10} & \multicolumn{1}{r|}{ \textbf{12.14}} & \multicolumn{1}{r}{ 10} \\
        & \multicolumn{1}{r|}{ MA } & \multicolumn{1}{r|}{ \textbf{1.55}} & \multicolumn{1}{r|}{ 10 }& \multicolumn{1}{r|}{ \textbf{10.45}} & \multicolumn{1}{r|}{ 10}& \multicolumn{1}{r|}{ \textbf{-30.51}} & \multicolumn{1}{r|}{ 20} & \multicolumn{1}{r|}{ \textbf{-6.74}} & \multicolumn{1}{r|}{ 5  } & \multicolumn{1}{r|}{ \textbf{4.29}} & \multicolumn{1}{r|}{ 10 }  & \multicolumn{1}{r|}{ \textbf{-9.91}} & \multicolumn{1}{r|}{ 10 } & \multicolumn{1}{r|}{ \textbf{11.61}} & \multicolumn{1}{r}{ 10} \\
        & \multicolumn{1}{r|}{ AU } & \multicolumn{1}{r|}{ -0.15} & \multicolumn{1}{r|}{ 0     } & \multicolumn{1}{r|}{ \textbf{4.12}} & \multicolumn{1}{r|}{ 50 } & \multicolumn{1}{r|}{ \textbf{-46.12}} & \multicolumn{1}{r|}{ 25} & \multicolumn{1}{r|}{ \textbf{-4.52}} & \multicolumn{1}{r|}{ 10 } & \multicolumn{1}{r|}{ 1.75\textsuperscript{\#}} & \multicolumn{1}{r|}{ 55  } & \multicolumn{1}{r|}{ \textbf{-21.44}} & \multicolumn{1}{r|}{ 30} & \multicolumn{1}{r|}{ 4.56\textsuperscript{\S}} & \multicolumn{1}{r}{ 0}   \\
        & \multicolumn{1}{r|}{ SU } & \multicolumn{1}{r|}{ \textbf{1.78}} & \multicolumn{1}{r|}{ 20 } & \multicolumn{1}{r|}{ \textbf{4.28}} & \multicolumn{1}{r|}{ 20 } & \multicolumn{1}{r|}{ \textbf{-19.06}} & \multicolumn{1}{r|}{ 30} & \multicolumn{1}{r|}{ \textbf{-3.04}} & \multicolumn{1}{r|}{ 15 } & \multicolumn{1}{r|}{ \textbf{1.69}} & \multicolumn{1}{r|}{ 15 }  & \multicolumn{1}{r|}{ \textbf{-7.01}} & \multicolumn{1}{r|}{ 40 } & \multicolumn{1}{r|}{ \textbf{17.31}} & \multicolumn{1}{r}{ 20} \\
        & \multicolumn{1}{r|}{ PH } & \multicolumn{1}{r|}{ \textbf{-1.13}} & \multicolumn{1}{r|}{ 20}  & \multicolumn{1}{r|}{ -0.33} & \multicolumn{1}{r|}{ 0     } & \multicolumn{1}{r|}{ \textbf{-4.09}} & \multicolumn{1}{r|}{ 15 } & \multicolumn{1}{r|}{ \textbf{-0.59}} & \multicolumn{1}{r|}{ 10 } & \multicolumn{1}{r|}{ \textbf{-1.37}} & \multicolumn{1}{r|}{ 15} & \multicolumn{1}{r|}{ \textbf{-2.18}} & \multicolumn{1}{r|}{ 15 } & \multicolumn{1}{r|}{ \textbf{-4.61}} & \multicolumn{1}{r}{ 20} \\
        & \multicolumn{1}{r|}{ BD } & \multicolumn{1}{r|}{ 0.77} & \multicolumn{1}{r|}{ 0      } & \multicolumn{1}{r|}{ 0.86} & \multicolumn{1}{r|}{ 0      } & \multicolumn{1}{r|}{ 0.44} & \multicolumn{1}{r|}{ 0       } & \multicolumn{1}{r|}{-0.22\textsuperscript{\S}} & \multicolumn{1}{r|}{ 0   } & \multicolumn{1}{r|}{ 1.85} & \multicolumn{1}{r|}{ 0      } & \multicolumn{1}{r|}{ 0.67} & \multicolumn{1}{r|}{ 0       } & \multicolumn{1}{r|}{ 17.43} & \multicolumn{1}{r}{ 0 }    \\
        & \multicolumn{1}{r|}{ AR } & \multicolumn{1}{r|}{ 0.16} & \multicolumn{1}{r|}{ 0      } & \multicolumn{1}{r|}{ -0.62\textsuperscript{\S}} & \multicolumn{1}{r|}{ 0  } & \multicolumn{1}{r|}{ \textbf{11.54}} & \multicolumn{1}{r|}{ 2 } & \multicolumn{1}{r|}{ \textbf{-1.19}} & \multicolumn{1}{r|}{ 1.5} & \multicolumn{1}{r|}{ 0} & \multicolumn{1}{r|}{ 0         } & \multicolumn{1}{r|}{ \textbf{5.6}} & \multicolumn{1}{r|}{ 2    }  & \multicolumn{1}{r|}{ 2.05} & \multicolumn{1}{r}{ 0}      \\
  \hline
    \end{tabular}%
     \begin{tablenotes}
    \scriptsize
    \item 1. \textbf{Performance Measures:} ACC- Accuracy, PRC- Precision, RCL- Recall, BS- Brier Score, F-M- F-Measure
    \item 2. \textbf{Datasets:} SO- Stack Overflow, MA- Mathematics, AU- Ask Ubuntu, SU- Super User, PH- Patch, BD- Bug-delay, AR- App-rating
    \item 3. Mag- Magnitude of the performance impact \textbar $x$-  \% of data of the noisy area when dropped starts statistically impacting the given performance measure
    \item 4.\textbf{ Cohen's d effect size:} Negligible - No formatting, Small -\textsuperscript{\S}, Medium -\textsuperscript{\#}, Large - \textbf{bold}
    \item 5. '$-$` indicates performance measure decreases due to removal of noisy area; '$+$` indicates performance measure increases due to removal of noisy area
    \item 6. All the values with small, medium or large effect size are statistically significant with $p \le 0.05$
    \item 7.  '$-$`in cases of Brier score indicates an actual increase in the Brier score and  '$+$` a decrease in Brier score (Lower the Brier score, the lesser the error)

\end{tablenotes}
\end{threeparttable}
  \label{tab:RQ1_results}
\end{table*}%

\noindent\textbf{Approach:} We employ our proposed framework (see Section~\ref{sec:framework}) to perform an incremental analysis as mentioned in Section~\ref{sec:deployment} to estimate the impact of discretization noise on the performance of a chosen classifier. We specifically draw attention to the classifier construction (step 3) of the framework (see Figure~\ref{fig:fig1}). In order to ascertain the performance impact of discretization noise on the classifiers, we train the chosen classifier on data after excluding incremental amounts of discretization noise. More specifically, we discard data points in windows which are defined as $cutpoint \pm cutpoint*x/100$ and use the retained data as the training data to build a classifier, where $x$ varies from $0$ to \textit{limit} in increments of $step\_size$ (as mentioned in Step 2 of our framework (See Section~\ref{sec:discretization})). The $step\_size$ can be different for different datasets depending on the \textit{limit} used to define the noise area for a particular dataset. Table~\ref{tab:limits} presents the various limit and \textit{step \_ size} values used for different datasets in our study. We perform this incremental analysis (see Fig 1 of Appendix) on all the studied datasets for the three different discretization thresholds (MT, CT, and, RTT) considered in the study as given in Section~\ref{sec:discretization} for all the four chosen classifiers (RFCM, LR, CART, and, KNN).

To measure whether the performance of a chosen classifier is impacted by removing data points in the noisy area (data containing discretization noise), we use a Wilcoxon signed-rank test~\cite{wilcoxon1945individual}, since it is a non-parametric test without any assumptions about the underlying distribution. Furthermore, to quantify the magnitude of the performance differences between the performance of the classifier with no data points removed and the classifier with data points in the noisy area removed, we use Cohen's $d$ effect size test~\cite{cohen1988statistical}. The threshold for analyzing the magnitude is as follows: $|d|$ $\leq$ 0.2 means magnitude is negligible,  $|d|$ $\leq 0.5$ means small, $|d|$ $\leq 0.8$ means medium and $|d|$ $>$ 0.8 means large.




We perform these statistical tests between the performance measures of the classifier that is constructed on the whole data and the classifier constructed on each step of the incremental analysis (where noisy points in the noisy area is incrementally removed). We do so to estimate how much data needs to be discarded to observe a statistically significant impact on the performance of a classifier. The $x$ value (of the $cutpoint \pm cutpoint*x/100$ used for discarding data in the noisy area) for which different performance measures get significantly impacted is reported. If discarding the whole of the noisy area does not create a statistically significant impact for a particular performance measure then 0 is reported instead of $x$ to signify that the discretization noise does not have a significant impact on that particular performance measure for the studied classifier.

\noindent\textbf{Results:} \textbf{The impact of discretization noise on the performance of different classifiers varies across datasets.} Similar performance impacts could be observed for the discretization noise generated by all the discretization thresholds considered in the study. Therefore, we only report the impact of discretization noise generated by the median based discretization threshold (MT) on various performance measures for all the four classifiers in Table~\ref{tab:RQ1_results} for brevity and space constraints (See Table~1 and Table 2 of the appendix for the performance impact due to other discretization thresholds on the studied classifiers).


Table~\ref{tab:RQ1_results} shows that the discretization noise impacts different classifiers differently. For instance, in the case of both CART and KNN classifiers, for all the datasets except for the patch dataset, the removal of discretization noise improves the performance in terms of AUC. However, for the patch dataset removal of discretization noise negatively impacts the AUC measure. For the same classifiers, even while the AUC and the Precision measures are positively impacted by the removal of discretization noise for (5/7 for CART and 6/7 for KNN), the Recall measure is negatively impacted for 5/7 datasets. Similarly, for LR classifier, while we observe no large impact on the AUC, we could observe up to 139\% impact on the Recall as we can observe from Table~\ref{tab:RQ1_results}. Furthermore, while the removal of discretization noise negatively impacts the accuracy of the RFCM and LR classifier, for the Stack Overflow dataset, it positively impacts the accuracy of the CART and KNN classifiers. Such a varied impact on the different performance measures for the different classifiers can be observed throughout (see Table~\ref{tab:RQ1_results}). \textbf{Therefore we assert that different classifiers are impacted differently (either positively or negatively) on the studied performance measures and datasets.}


Additionally, in Table~\ref{tab:RQ1_results} we also provide the $x$, which tells us the percentage of data from the noisy area, that when dropped, starts statistically impacting the given performance measure. This value aids the users of our framework to know how much data they need to discard in order to avoid the performance impact of the noise on a particular performance measure for a studied classifier.

We find that the removal of discretization noise has both a positive and negative impact on different performance measures for different classifiers. In addition, we do not observe a generalizable trend in how the discretization noise affects different performance measures. For instance, from Table~\ref{tab:RQ1_results}, we see that the removal of the noisy area from datasets negatively impacts the performance measures of an RFCM for 6/7 datasets on Accuracy, Recall, Brier score, F-measure, 2/7 datasets on AUC, and 5/7 datasets on MCC, most of which in a statistically significant fashion with and a large effect size. However, for 6/7 datasets, removal of noisy area positively impacts the Precision in a statistically significant fashion with a large effect size (in most cases). Furthermore, while removal of noisy area negatively impacts the performance measures for most datasets, it improves the Accuracy, Precision, Recall, F-Measure, and MCC for the App-rating dataset in a statistically significant fashion with a large effect size. These varied impacts of the discretization noise on the different performance measures highlight the need for our frameworks to study the peculiarities of each case study as \textbf{the discretization noise affects the different performance measures differently}

Finally, we also note that \textbf{magnitude of the performance impact due to discretization noise varies for different performance measures.} For instance, in all of the classifiers, for most of the datasets, we could see that while Recall is impacted heavily (up to 139\% for LR in Ask Ubuntu), other performance measures even if impacted significantly, are not at the same magnitude (e.g, only -5.54\% for accuracy in LR for Ask Ubuntu). \\

\textbf{Discussion:}
We find that the magnitude of the impact of some performance measures is much greater than other performance measures. Also, from Table~\ref{tab:RQ1_results} we note that for a given dataset and a classifier, some performance measures are impacted even for small amounts of discretization noise (given by the $x$ in the Table~\ref{tab:RQ1_results}). Whereas, some other performance measures are more resilient. For instance, in the case of the Mathematics dataset for the RFCM classifier, discarding 15\% of the data in the noisy area significantly impacts both Precision and Recall with a large magnitude varying from 8.96\% to -32.2\%. Whereas even with a drop of 70\% data from the noisy area the AUC gets impacted marginally by -1.23\%, which indicates that \textbf{some performance measures are more resistant to discretization noise than others}.

The different degrees to which different performance measures are impacted can be attributed to the different nature of each performance measure and what they seek to capture. For instance, Precision, Recall, and F-measure focus on capturing how good a classifier performs in predicting one of the classes~\cite{sokolova2006beyond}. Therefore, a potential imbalance in the dataset caused by discarding different amounts of data in the noisy area, even if it is discretization noise (as such noisy points contain useful information too), could impact these measures greatly. Furthermore, Flach shows that these measures are easily impacted by class imbalance~\cite{flach2003geometry}. These explain the large impacts that we observe for the Precision, Recall, and F-measures (as opposed to Accuracy which takes both the classes into consideration). Especially, with all of our datasets having a skewed distribution of the dependent variable (average Skewness of the dependent variable of all of the studied datasets is 12.99 and average Kurtosis is 309.4). For example, let us consider the Ask Ubuntu dataset with MT as the discretization threshold. When the whole dataset is used (without any removal of data), the number of data points belonging to each class are equal. However, discarding the data in the noisy area according to Table~\ref{tab:limits} for MT induces a class imbalance in the dataset. The number of data points belonging to ``class1" only makes up 35\% of the dataset. Such class imbalance produced by discarding data induces a high degree of impact on class-specific performance measures for the Ask Ubuntu dataset as shown in Table~\ref{tab:RQ1_results}.

However, the more balanced measures like AUC and MCC are more robust and insensitive to class distributions~\cite{lessmann2008benchmarking}. Therefore, AUC and MCC are impacted much less severely by the discretization noise in the dataset. As we can observe from Table~\ref{tab:RQ1_results}, the magnitude to which measures like AUC and MCC are impacted is lesser than the class-specific measures like Precision, Recall, and F-measure. For instance, while Recall is impacted by as much as 139\%, AUC is impacted only by at most 4.29\% across all the datasets and classifiers. Even for the Ask Ubuntu dataset, while the impact on class-specific performance measures is high, the AUC is seldom impacted. A similar trend could be observed for MCC as they consider both false positives and false negatives, even though they are slightly more sensitive to the discretization noise than AUC. As Huang~\etal~\cite{huang2005using} and Mossman~\cite{mossman1994assessing} show balanced measures like AUC are very stable and insensitive to noise hence even a small impact in terms of magnitude (if the effect size is large) could be significant. Therefore though the absolute magnitude of the impact is small for some performance measures like AUC and Accuracy, the true nature of the impact needs to established by the practitioner. Our framework seeks to provide a means for finding and measuring the impact.  Finally, similar to other balanced measures we observe that the impact on Brier score is moderate (within 13\%) when compared to class-specific performance measures. Brier score being an error metric calculates the mean squared difference between the actual outcome and the assigned probability, which elucidates the distance between the classifier's predictions and the actual classes in probability scale. Table~\ref{tab:RQ1_results} shows that Brier score is universally negatively impacted for a RFCM (signifying an increase in Brier score). Because, though the noisy points contain discretization noise, they also contain useful information (see Section~\ref{sec:disc1}) and the removal of such information could universally affect the predicted probability scores of classifiers, especially when they are robust to noise~\cite{folleco2008software}. However the Table~1 and Table~2 provided in Appendix C (for performance impact due to discretization noise that is generated by CTT and RT) shows a varied but moderate impact (within 14\%) for LR, CART and KNN, which is similar to other balanced performance measures. \textbf{In summary, unlike the class-specific performance measures, balanced measures are impacted moderately by the discretization noise.}


To conclude, discretization noise impacts different performance measures differently for various datasets with varying magnitudes for different classifiers (0-139\%) as shown in Table~\ref{tab:RQ1_results}. We also note that \textbf{balanced performance evaluation measures are more resilient to the discretization noise whereas class-specific performance evaluation measures are greatly impacted by the discretization noise in the dataset}. Thus the inclusion of discretization noise in the construction of any chosen classifier could be either beneficial or detrimental depending on the performance measure of interest and the dataset at hand. \textbf{Such unpredictability demonstrates the need for our framework to better understand when it is advisable to remove the noisy points, and how much of the data in the noisy area is to be removed to avoid impact on the studied performance measure for a chosen classifier.}\\

\vspace{-0.33cm}
\rqbox{The impact of noisy discretization data points is inconsistent across multiple performance measures for different datasets and different. Though the impact on class-specific measures (Precision, Recall,F-measure) is the most pronounced (up to 139\%) other performance measures are also consistently impacted at least up to 60\% due to the inclusion or exclusion of discretization noise (both positively and negatively). Hence, it is advisable to use our framework beforehand to carefully understand if the noisy discretization data points are to be used or discarded and how much of them needs to be discarded.}
\vspace{0.2cm}

\vspace{-1cm}
\subsection{\textbf{Studying the Impact of Discretization Noise on the Interpretation of a Classifier}}\label{sec:rq2}
\noindent\textbf{Motivation:} Many prior studies use classifiers to understand the impact of features on the dependent variable~\cite{mcintosh2016empirical,thongtanunam2016revisiting}. From Section~\ref{sec:rq1}, we observe that the data from the noisy area sometimes impacts the performance of a classifier differently for different datasets. This could be because the data in the noisy area contains useful information along with the discretization noise. Therefore removing/including such data may also lead to a misleading interpretation of a classifier. Therefore, we seek to observe how the discretization noise impacts the derived feature importance with our proposed framework to decide if such noisy data points can be safely discarded or should they be included nevertheless.


\noindent\textbf{Approach:} We demonstrate the capability of our framework to analyse the impact of discretization noise on the derived feature importance ranks. We adopt the same incremental analysis approach that we adopted in the Section~\ref{sec:rq1}. But instead of measuring the performance of the classifiers that are constructed by incrementally discarding data from the noisy area, we note the feature importance values of these classifiers (see step 3 of Section~\ref{sec:framework}). We perform an incremental analysis on all the studied datasets for the three different discretization thresholds (MT, CT, RTT) considered in the study in Section~\ref{sec:discretization} (see Fig~1 of Appendix) for all the four chosen classifiers (RFCM, LR, CART, KNN).


We measure the derived feature importance values of the chosen classifier trained on various data configurations. We use the Scott-Knott ESD test to rank~\cite{tantithamthavorn2017empirical,tantithamthavorn2016automated,li2017towards,Ghotra:2015:RIC:2818754.2818850} the features with their feature importance values. To observe whether the derived feature importance ranks vary significantly, we compare the derived feature importance ranks of the classifier that is trained on the whole data and the classifier with $x=limit$ (see Table~\ref{tab:limits}) data points removed from the noisy area for each of studied dataset across all the discretization thresholds. We compute the difference between the derived feature importance ranks for each feature in the dataset and compare them to the null distribution (where the rank difference of each feature is zero) to see if removing the noisy area impacts the derived feature importance ranks of a classifier~\cite{rajbahadur2017impact}.

For instance, lets consider the Stack Overflow dataset with MT as the discretization threshold when used for training an RFCM. We wish to study the impact of discretization noise on the interpretation of the constructed RFCM. For simplicity, lets consider that the Stack Overflow dataset has only 5 features. For the RFCM that is intially trained on the whole dataset, a derived feature importance ranks is generated for its 5 features ($F\textsubscript{whole}= {3,1,5,4,2}$). Following which, an RFCM is trained on the Stack Overflow dataset devoid of the noisy area
($F\textsubscript{noisy\_area\_removed}= {2,1,3,2,4}$). The difference between $F\textsubscript{whole}$ and $F\textsubscript{noisy\_area\_removed}$
is \textit{difference} $= {1,0,2,2,2}$. If the $F\textsubscript{whole}$ and $F\textsubscript{noisy\_area\_removed}$ are the same, then the \textit{difference }generated would be zero and would imply that the discretization noise does not impact the interpretation of a classifier. If the difference is not 0 (as in our example), we compare the \textit{difference} against a zero distribution (\textit{null\_ distribution}$={0,0,0,0,0}$) with a Wilcoxon-signed rank test and Cohen's effect size test to determine whether if the difference is significant or otherwise.

\begin{figure*}
    \includegraphics[width=\linewidth]{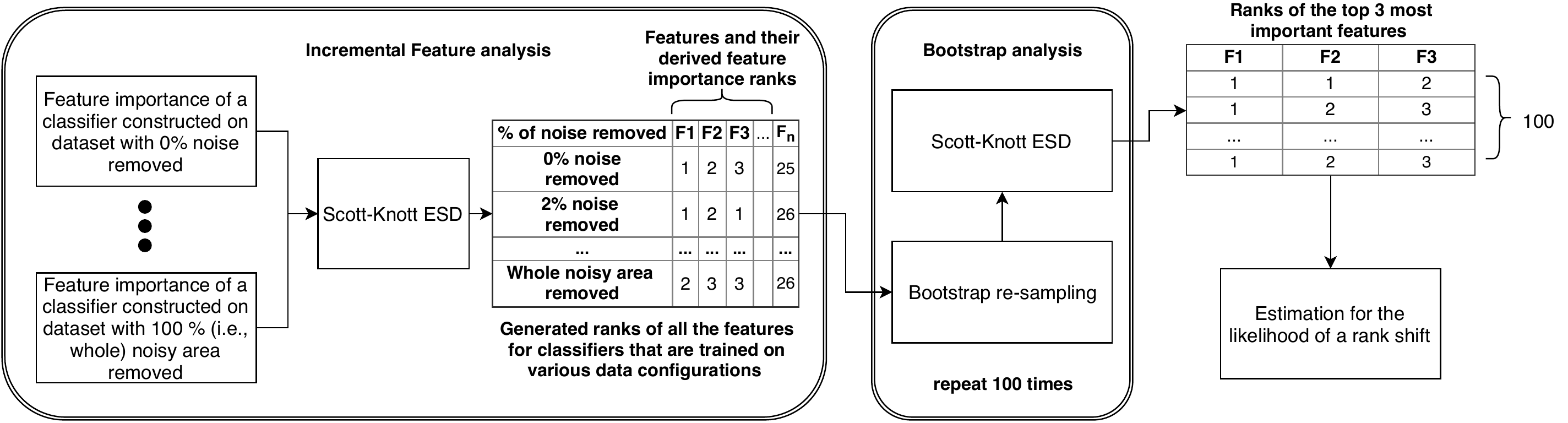}

    \caption{The procedure for estimation of the likelihood of a rank shift.}

    \label{fig:rq2_2}
\end{figure*}




In addition, even if the discretization noise impacts the overall interpretation of a classifier, most researchers and practitioners care only about the top x most important features~\cite{hassan2005top,lewis2013does}. Therefore, in this section we present the results of the top 3 features as a showcase. To measure how likely the rank of a feature could shift due to the removal of the discretization noise, we compute the likelihood of rank shifts for top 3 ranks. The importance rank of a feature is ascertained by the median rank of the derived feature importance ranks for a feature of the dataset.

We use a bootstrap analysis to compute the likelihood of rank shifts similar to prior study~\cite{tantithamthavorn2016automated}. Figure~\ref{fig:rq2_2} outlines the process of estimation of the likelihood of a rank shift. The feature importance ranks that are generated from each classifier in the incremental feature analysis are taken as the input. These derived ranks are then re-sampled with replacement (i.e., Bootstrap re-sampling). The bootstrap re-sampled feature rank distribution is generated for all the features of each of the studied datasets and they are re-ranked using the Scott-Knott ESD test as shown in the ``Bootstrap analysis'' part of Figure~\ref{fig:rq2_2}. This process is repeated 100 times. The intuition behind such a procedure is that the bootstrap re-sampling and re-ranking would alleviate possible minor and insignificant fluctuations in the derived feature importance ranks while highlighting the pattern of significant fluctuations in the derived feature importance ranks for a feature, and thereby bringing out its true rank. Now we have 100 derived feature importance ranks for all the features in each of the studied datasets.



These 100 ranks (the output of the ``Bootstrap analysis'' part of Figure~\ref{fig:rq2_2}) for a feature are used for estimating the likelihood of a rank shift. The likelihood of a rank shift for rank $x$ feature is computed as the percentage of how many ranks are not equal to $x$. For example, for rank 1 feature, out of 100 times 2 ranks are not equal to 1, then the likelihood of a rank shift for that feature is 0.02. The estimation for the likelihood of rank shifts is done for the features in the top 3 ranks of all the studied datasets. If the likelihood values are high for a particular rank in a dataset, it indicates that discretization noise impacts the derived feature importance and the feature(s) reported at that rank should be interpreted with caution in that dataset. 

\begin{table}[]
\scriptsize
\centering
\caption{The likelihood of rank shifts in the top 3  most important ranks (column A) and the comparison of the derived feature importance ranks of an RFCM trained on the whole dataset (Rank$_{W}$) and the dataset with the noisy area removed (Rank$_{NR}$) (column B).}
\label{tab:likelihood_table}
\begin{tabular}{l|p{0.1cm}|p{0.1cm}|p{0.1cm}|p{0.1cm}|p{1.1cm}}
\hline
\multirow{2}{*}{\textbf{Dataset}} & \multicolumn{3}{c|}{\textbf{Rank shift likelihood (A)}}&  \multicolumn{2}{c}{\textbf{Rank$_W$ vs. Rank$_{NR}$ (B)}}                                          \\ \cline{2-6}
                                  & \multicolumn{1}{l|}{\textbf{Rank 1}} & \multicolumn{1}{l|}{\textbf{Rank 2}} & \multicolumn{1}{l|}{\textbf{Rank 3}} & \multicolumn{1}{l|}{\textbf{p-value}} & \multicolumn{1}{l}{\textbf{Cohen's d}}

                                  \\ \hline
Stack Overflow                    & 0                                    & 0                                & 0  & 0                                     & -1.29 (L)                                  \\
Mathematics                       & 0                                    & 0                                & 0 &  0                                     & -2.14 (L)                                        \\
Ask Ubuntu                        & 0                                    & 0                                & 0   & 0                                     & -2.84 (L)                               \\
Super User                        & 0                                    & 0                                & 0  & 0.01                                  & -0.74 (M)                                 \\
Patch                             & 0                                    & 0                                 & 0  & 0.03                                  & -0.62 (M)                                \\
Bug-delay                         & 0                                    & 0                               & 0   & 0                                  & -1.39 (L)                               \\
App-rating                        & 0                                    & 0                              & 0   & 1                                     & -0.30 (S)                                 \\
\hline
\end{tabular}
\end{table}

\noindent\textbf{Results:} \textbf{The overall derived feature importance ranks are impacted by the discretization noise for most of the studied datasets.} We only report the impact of discretization noise generated by median based discretization threshold (MT) on the derived feature importance ranks for RFCM due to space constraints, However similar results are noted on all the other classifiers (LR, CART, KNN) on all the studies discretization thresholds (please refer to Table~3, 4, and 5 of Appendix C for other results). We show a comparison of the derived feature importance ranks of an RFCM that is trained on the dataset with and without the noisy area removed in Table~\ref{tab:likelihood_table}. The results highlight that for most of the studied datasets (6/7 datasets in the case of the RFCM classifier and for all the datasets for other classifiers), the overall derived feature importance ranks of all the features in a dataset, are significantly impacted (i.e., $p-$value $<$ 0.05) with a non-negligible effect size.




\textbf{However, the derived feature importance ranks of the top 3 features are not impacted by the discretization noise.} To specifically understand how much of an impact that the discretization noise has on the most important features, we computed the likelihood of a rank shift for the top 3 important features. We present the results in Table~\ref{tab:likelihood_table}. \textbf{For all the datasets, the ranks of the most important features (i.e., all features at rank 1, rank 2 and rank 3) are not impacted due to discretization noise}. We further observe that these trends are not specific to RFCM and the discretization noise generated by MT. From Table 3, 4, and 5 (please see Appendix C) we observe that the most important features are not impacted by the discretization noise generated in the dataset for any of the studied classifiers.

In summary, though the overall ranks of derived feature importance are impacted by the discretization noise in the dataset, the top 3 (most important) features that most researchers and practitioners focus on~\cite{hassan2005top,lewis2013does,rajbahadur2017impact,tantithamthavorn2015icse} are not impacted by the discretization noise. \textbf{Therefore, we suggest that the decision of either including or removing the data in the noisy area could be exclusively arrived at from the results of Section~\ref{sec:rq1} without being worried much about its impact on interpretation}. Nevertheless these results might vary for other settings (e.g., other datasets or classifiers) and our framework is able to provide a case by case guidance.




\rqbox{The discretization noise (generated by our studied discretization thresholds) does not impact the derived feature importance of any of the top 3 features yet it impacts the overall derived feature importance ranks. }
\section{Discussion}\label{sec:disc}

\begin{table*}
    \centering
    \scriptsize
    \caption{Median performance (AUC) of the RFCMs on the different regions of the data.\label{tab:prem_tab_2}}
    \begin{tabular}{l|lllllll}
        \hline
        \textbf{Training data $\rightarrow$ Testing data}& \textbf{Stack Overflow}& \textbf{Mathematics}& \textbf{Ask Ubuntu}& \textbf{Super User}& \textbf{Patch}& \textbf{Bug-delay}& \textbf{App-rating}\\
        \hline
        \textbf{Noisy area $\rightarrow$ Extremes} & \multicolumn{1}{r}{0.96} & \multicolumn{1}{r}{0.96}& \multicolumn{1}{r}{0.86}& \multicolumn{1}{r}{0.86} & \multicolumn{1}{r}{0.98}& \multicolumn{1}{r}{0.69}& \multicolumn{1}{r}{0.76}\\

        \hline


        \hline
    \end{tabular}

\end{table*}
\subsection{Why does a classifier trained on the whole dataset (with discretization noise) sometimes perform better than the classifier trained on data devoid of discretization noise?}
From the Section~\ref{sec:rqs}, we observe that excluding data from the noisy area, i.e., data with discretization noise, sometimes negatively impacts the performance of a classifier.  We seek to understand this counter intuitive phenomena. Furthermore, in this section, we also remark about why the top 3 important features of a classifier are not impacted by discretization noise.

We hypothesize that a classifier in some cases is able to capture the signal from the noisy points in spite of the discretization noise. Doing so, allows the classifier to capture more information from the noisy data points, in addition to information available from the clean data. Therefore, discarding those noisy points negatively impacts the performance of a classifier. To examine our hypothesis, we start by constructing classifiers with data points from the noisy area (generated with MT) and test them on the data from the extremes and noisy areas. Such an experiment helps us understand whether the data with the discretization noise contains any useful information.

We follow the steps outlined in our framework (see Section~\ref{sec:framework}) to construct the classifiers on the noisy area and generate the out-of-sample test sets from the extremes and noisy area separately. For this experiment, we construct an RFCM and observe its performance on the AUC measure. We do so just on RFCM as our intention is only to analyse if the noisy area contains useful information and our results on RFCM could help us test it succinctly. Furthermore,  in this section we report only the AUC measure as from Table~\ref{tab:RQ1_results} we could observe that AUC measure is the most resilient performance measure that has the least impact across all the studied classifiers and we wish to report the impact on the most resilient measure. A high AUC would therfore objectively justify the presence of useful information. However, all the other performance measures follow the same trend.

\textbf{The noisy area contains useful information that might help improve the performance of RFCMs.} From Table~\ref{tab:prem_tab_2}, we observe that the RFCMs that are trained on the noisy area perform extremely well on the extremes for 5 out of the 7 studied datasets. Three datasets have an AUC that is larger than 0.95. For instance, in the Stack Overflow dataset, the RFCMs that are trained on the noisy area have an AUC of 0.96 when tested on the extreme areas.

While the previous experiments show that noisy area contains useful information, we cannot conclusively establish if the contained information in the noisy points amidst the discretization noise could be successfully used by the classifiers. To test if the studied classifiers can use the information contained in the noisy area, we construct classifiers that are trained with extremes and data from the noisy area (in contrary to the previous experiment, where we trained only on the noisy area) and then add increasing amounts of data from the noisy area. If the performance of a classifier does not degrade significantly (some degradation should be expected) with the increased amount of noise, it may indicate that the classifier is capable of capturing a signal as long as there is enough information in the data. Therefore, we could infer that despite the presence of discretization noise, the data points in the noisy area provide an additional signal to the classifier and the exclusion of the noisy area, negatively impacts the performance of the classifier. On the other hand, if there is a drastic and significant performance degradation of the classifier, it would then invalidate our hypothesis that classifiers are able to learn an additional signal in spite of discretization noise. 




We perform the experiment by setting up a simulation study with the help of our framework similar to the previous experiment, where we train all the classifiers on the extremes with different amount data from the noisy area. We train our classifiers on four different data configurations which are given by $(extremes + (noisy area + over\_sample\% *(noisy area))$, where the $over\_sample$ takes values of $0,100,200,300$. We oversample different amounts of data from the noisy area while keeping the amount of the data from extremes constant.  We build the classifiers on four data configurations i.e., $(extremes + (noisy  area + 0\% *(noisy area))$, $(extremes + (noisy area+ 100\% *(noisy\_area))$, $(extremes + (noisy area + 200\% *(noisy\ area))$, and $(extremes + (noisy area +300\% *(noisy\ area))$. For instance, in the Ask Ubuntu dataset from our study for MT, the extremes have 1,427 data points and the noisy area has 2,711 data points as shown in the Table~\ref{tab:limits}. Therefore for the first configuration, we would have 4,138 data points, of which 65\% is comprised of noisy points. Therefore for Ask Ubuntu dataset, our four data configuration consist of 65\%, 79\%, 85\% and 88\% noisy points respectively along with the clean data from the extremes. (See Figure~3 of Appendix~C explaining the overall experimental setup)

 The performance of a classifier that is constructed on the aforementioned data configurations is evaluated on the out-of-sample test data from the extremes. Note that the out-of-sample test data that is obtained from the extremes, is not used in the training phase and is only used for testing the constructed classifier. The experimental setup for constructing a classifier is similar to that of our framework, as outlined in Section~\ref{sec:framework}. Finally, we also capture the derived feature importance ranks and observe the likelihood of rank shifts for the top 3 most important features as outlined in Section~\ref{sec:framework} and Section~\ref{sec:rq2}.

\textbf{Adding data from the noisy area to the training data does not greatly impact the performance of a classifier.} We report the results in Table~\ref{tab:Disc1_tab1}. The columns in Table~\ref{tab:Disc1_tab1} correspond to different data configurations that we discussed earlier. We observe that the median AUC of a classifier that is trained on the dataset without noise is quite close to the AUC of a classifier that is trained on data with 300\% noise. For RFCM, LR and CART classifiers, even an addition of 300\% of data from the noisy area only impacts the AUC within 6\% as we can observe from Table~\ref{tab:Disc1_tab1}. Even in the case of the KNN classifier, which is an instance-based classifier that is traditionally more sensitive to noise in the data~\cite{aha1989noise}, gets impacted only by 11\% in terms of AUC even with the addition of up to 300\% data points from the noisy area. These results signify that the performance of the constructed classifiers on the extremes does not degrade significantly even when there is $300\%$ (at least X\% of the data is noisy) data from the noisy area in addition to data from the extremes.

Furthermore, we also note that the likelihood of rank shifts for top 3 ranks between classifiers that are trained on the first configuration (0\%) and the last configuration (300\%) is 0. Which further reinforces the validity of our hypothesis that the classifiers are able to capture the signal in the noisy points despite the discretization noise and the most important features contributed by the true signal in the underlying data are not perturbed by the discretization noise.

\begin{table}[htbp]
  \begin{threeparttable}
      \caption{Performance comparison (in AUC) of classifiers that are trained on different data configurations.}\label{tab:Disc1_tab1}
\scriptsize
    \begin{tabular}{p{1cm}|p{0.8cm}|p{1cm}|p{1cm}|p{1cm}|p{1cm}}
     \hline
    \textbf{Classifier} & \textbf{Dataset} & \textbf{0\% Noise} & \textbf{100\% Noise} & \textbf{200\% Noise} & \textbf{300\% Noise} \\
    \hline
    \multirow{7}[2]{*}{RF} & \multicolumn{1}{l|}{ SO} & \multicolumn{1}{r|}{ 0.96} & \multicolumn{1}{r|}{ 0.96} & \multicolumn{1}{r|}{ 0.96} & \multicolumn{1}{r}{ 0.96} \\
       & \multicolumn{1}{l|}{ MA} & \multicolumn{1}{r|}{ 0.96} & \multicolumn{1}{r|}{ 0.96} & \multicolumn{1}{r|}{ 0.96} & \multicolumn{1}{r}{ 0.96} \\
       & \multicolumn{1}{l|}{ AU} & \multicolumn{1}{r|}{ 0.86} & \multicolumn{1}{r|}{ 0.85} & \multicolumn{1}{r|}{ 0.84} & \multicolumn{1}{r}{ 0.84} \\
       & \multicolumn{1}{l|}{ SU} & \multicolumn{1}{r|}{ 0.86} & \multicolumn{1}{r|}{ 0.86} & \multicolumn{1}{r|}{ 0.87} & \multicolumn{1}{r}{ 0.85} \\
       & \multicolumn{1}{l|}{ PT} & \multicolumn{1}{r|}{ 0.99} & \multicolumn{1}{r|}{ 0.99} & \multicolumn{1}{r|}{ 0.99} & \multicolumn{1}{r}{ 0.99} \\
       & \multicolumn{1}{l|}{ BD} & \multicolumn{1}{r|}{ 0.69} & \multicolumn{1}{r|}{ 0.68} & \multicolumn{1}{r|}{ 0.66} & \multicolumn{1}{r}{ 0.65} \\
       & \multicolumn{1}{l|}{ AR} & \multicolumn{1}{r|}{ 0.76} & \multicolumn{1}{r|}{ 0.77} & \multicolumn{1}{r|}{ 0.76} & \multicolumn{1}{r}{ 0.76} \\
    \hline
    \multirow{7}[2]{*}{LR} & \multicolumn{1}{l|}{ SO} & \multicolumn{1}{r|}{ 0.92} & \multicolumn{1}{r|}{ 0.93} & \multicolumn{1}{r|}{ 0.92} & \multicolumn{1}{r}{ 0.92} \\
       & \multicolumn{1}{l|}{ MA} & \multicolumn{1}{r|}{ 0.93} & \multicolumn{1}{r|}{ 0.92} & \multicolumn{1}{r|}{ 0.92} & \multicolumn{1}{r}{ 0.92} \\
       & \multicolumn{1}{l|}{ AU} & \multicolumn{1}{r|}{ 0.78} & \multicolumn{1}{r|}{ 0.77} & \multicolumn{1}{r|}{ 0.76} & \multicolumn{1}{r}{ 0.76} \\
       & \multicolumn{1}{l|}{ SU} & \multicolumn{1}{r|}{ 0.79} & \multicolumn{1}{r|}{ 0.79} & \multicolumn{1}{r|}{ 0.78} & \multicolumn{1}{r}{ 0.77} \\
       & \multicolumn{1}{l|}{ PT} & \multicolumn{1}{r|}{ 0.98} & \multicolumn{1}{r|}{ 0.98} & \multicolumn{1}{r|}{ 0.98} & \multicolumn{1}{r}{ 0.97} \\
       & \multicolumn{1}{l|}{ BD} & \multicolumn{1}{r|}{ 0.68} & \multicolumn{1}{r|}{ 0.68} & \multicolumn{1}{r|}{ 0.66} & \multicolumn{1}{r}{ 0.66} \\
       & \multicolumn{1}{l|}{ AR} & \multicolumn{1}{r|}{ 0.72} & \multicolumn{1}{r|}{ 0.72} & \multicolumn{1}{r|}{ 0.71} & \multicolumn{1}{r}{ 0.71} \\
    \hline
    \multirow{7}[2]{*}{CART} & \multicolumn{1}{l|}{ SO} & \multicolumn{1}{r|}{ 0.89} & \multicolumn{1}{r|}{ 0.86} & \multicolumn{1}{r|}{ 0.84} & \multicolumn{1}{r}{ 0.83} \\
       & \multicolumn{1}{l|}{ MA} & \multicolumn{1}{r|}{ 0.87} & \multicolumn{1}{r|}{ 0.86} & \multicolumn{1}{r|}{ 0.87} & \multicolumn{1}{r}{ 0.85} \\
       & \multicolumn{1}{l|}{ AU} & \multicolumn{1}{r|}{ 0.74} & \multicolumn{1}{r|}{ 0.69} & \multicolumn{1}{r|}{ 0.67} & \multicolumn{1}{r}{ 0.66} \\
       & \multicolumn{1}{l|}{ SU} & \multicolumn{1}{r|}{ 0.72} & \multicolumn{1}{r|}{ 0.70} & \multicolumn{1}{r|}{ 0.69} & \multicolumn{1}{r}{ 0.68} \\
       & \multicolumn{1}{l|}{ PT} & \multicolumn{1}{r|}{ 0.94} & \multicolumn{1}{r|}{ 0.89} & \multicolumn{1}{r|}{ 0.90} & \multicolumn{1}{r}{ 0.90} \\
       & \multicolumn{1}{l|}{ BD} & \multicolumn{1}{r|}{ 0.64} & \multicolumn{1}{r|}{ 0.62} & \multicolumn{1}{r|}{ 0.60} & \multicolumn{1}{r}{ 0.60} \\
       & \multicolumn{1}{l|}{ AR} & \multicolumn{1}{r|}{ 0.64} & \multicolumn{1}{r|}{ 0.63} & \multicolumn{1}{r|}{ 0.62} & \multicolumn{1}{r}{ 0.62} \\
    \hline
    \multirow{7}[2]{*}{KNN} & \multicolumn{1}{l|}{ SO} & \multicolumn{1}{r|}{ 0.80} & \multicolumn{1}{r|}{ 0.75} & \multicolumn{1}{r|}{ 0.71} & \multicolumn{1}{r}{ 0.69} \\
       & \multicolumn{1}{l|}{ MA} & \multicolumn{1}{r|}{ 0.80} & \multicolumn{1}{r|}{ 0.75} & \multicolumn{1}{r|}{ 0.72} & \multicolumn{1}{r}{ 0.70} \\
       & \multicolumn{1}{l|}{ AU} & \multicolumn{1}{r|}{ 0.62} & \multicolumn{1}{r|}{ 0.61} & \multicolumn{1}{r|}{ 0.59} & \multicolumn{1}{r}{ 0.58} \\
       & \multicolumn{1}{l|}{ SU} & \multicolumn{1}{r|}{ 0.69} & \multicolumn{1}{r|}{ 0.64} & \multicolumn{1}{r|}{ 0.62} & \multicolumn{1}{r}{ 0.61} \\
       & \multicolumn{1}{l|}{ PT} & \multicolumn{1}{r|}{ 0.83} & \multicolumn{1}{r|}{ 0.82} & \multicolumn{1}{r|}{ 0.81} & \multicolumn{1}{r}{ 0.80} \\
       & \multicolumn{1}{l|}{ BD} & \multicolumn{1}{r|}{ 0.55} & \multicolumn{1}{r|}{ 0.52} & \multicolumn{1}{r|}{ 0.51} & \multicolumn{1}{r}{ 0.50} \\
       & \multicolumn{1}{l|}{ AR} & \multicolumn{1}{r|}{ 0.60} & \multicolumn{1}{r|}{ 0.57} & \multicolumn{1}{r|}{ 0.56} & \multicolumn{1}{r}{ 0.55} \\
    \hline
    \end{tabular}%
           \begin{tablenotes}
    \scriptsize

    \item \textbf{Datasets:} SO- Stack Overflow, MA- Mathematics, AU- Ask Ubuntu, SU- Super User, PH- Patch, BD- Bug-delay, AR- App-rating

\end{tablenotes}
    \end{threeparttable}

\end{table}%


In summary, We establish that the noisy area does contain some useful information. Further, we observe only a maximum performance drop of 11\% across 7 datasets for all of the studied classifiers (with less than 6\% performance drop for RFCM, LR and CART) with the addition of as much as $300\%$ of data from the noisy area, where at least more than 67\% of the dataset is noisy. This suggests that a classifier is able to capture the information in the data in spite of the noise, thereby explaining why the exclusion of data points from the noisy area sometimes impacts the performance of a classifier. In addition, the likelihood of rank shift for the top 3 most important features is 0 for all the classifiers, signifying that the discretization noise even in such high quantities does not impact the interpretation of the classifiers.

\label{sec:disc1}
\subsection{Why does inclusion of discretization noise sometimes negatively impacts the performance of a classifier?}

Contrary to Section~\ref{sec:disc1}, in this section, we seek to understand why the inclusion of discretization noise negatively impacts the performance of some classifiers. From Table~\ref{tab:RQ1_results} we observe that for all the studied classifiers, the inclusion of discretization noise sometimes negatively impacts the performance of a classifier even though Section~\ref{sec:disc1} shows that data in the noisy area has useful information and classifiers are capable of leveraging it. From Table~\ref{tab:RQ1_results} we also observe that for all the studied classifiers and datasets, at least one of the performance measure is negatively impacted. We hypothesize that such a negative impact could be due to the high complexity (less discriminative power) of the noisy points around the discretization threshold, despite containing useful information. We arrive at such a hypothesis as prior studies show that it is difficult for the classifiers to perform well if the complexity of the data is high, irrespective of the contained information~\cite{Alm:2005:ETM:1220575.1220648,ho2002complexity}. Thus, we are interested in exploring if the negative impact in the performance of a classifier due to the inclusion of discretization noise is because of the high complexity of the data points around the discretization threshold (noisy area).


\begin{figure}
    \includegraphics[width=\linewidth]{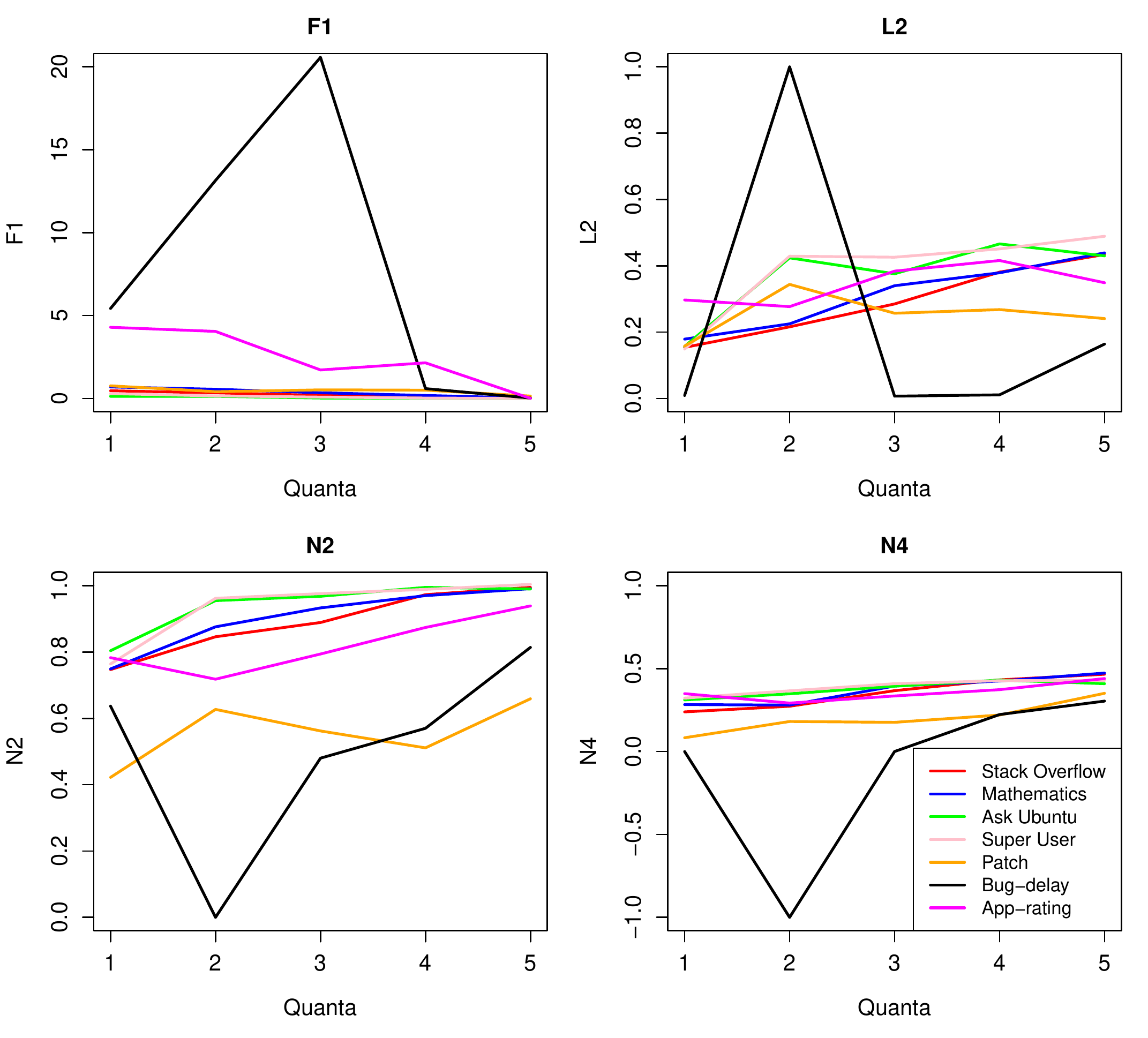}

    \caption{Data complexity across quantum for the studied datasets.}

    \label{fig:disc2_1}
\vspace{-0.4cm}
\end{figure}

Ho and Basu~\cite{ho2002complexity} provide complexity metrics to measure the complexity of data. We use these complexity metrics to measure the complexity of different regions of our data. From the multiple methods that are proposed by Ho and Basu~\cite{ho2002complexity}, we choose Fisher's discriminant ratio (F1), linear separability (L2), mixture identifiability (N2) and nonlinearity (N4) as they are simple to explain and easy to interpret. We briefly explain the metrics that we choose in the Table~6 of Appendix~C. See the study by Ho and Basu~\cite{ho2002complexity} for more details about the computation of these metrics.

We use the abovementioned measures to compute how the complexity of the dataset changes across data points as we move from the extremes to the noisy areas. We first discretize the data into ``class 1'' and ``class 2''  classes as outlined in Section~\ref{sec:discretization}. We then transform the continuous dependent variable using the Box-Cox transformation~\cite{sakia1992box} to alleviate the skew and increase the spread of the distribution of the dependent variable. We then split the data into 5 quanta for each class using the \textbf{bin} function in R. We do so to compartmentalize the data in relation to the continuous dependent variable and analyse the changes to the complexity of the data points as we move closer to the discretization threshold (we choose MT) for our case study. We would not be able to observe how the complexity changes in different areas of the data without such compartmentalization of the data. The choice of using 5 quanta is so that the compartmentalization is neither too granular nor too encompassing. The 1\textsuperscript{st}quantum contains most of the data from extremes and the 5\textsuperscript{th} quantum contains most of the data from the noisy area, whereas 2\textsuperscript{nd} to 4\textsuperscript{th} quantum roughly contain an equivalent amount of data points in between. Finally, we compute the above-mentioned data complexity metrics for the data points in each of these quantum and plot the results. (See Figure~2 of Appendix~C explaining the overall experimental setup)

From Figure~\ref{fig:disc2_1}, we observe that as we move from the extremes (1\textsuperscript{st} quantum) towards the noisy area (5\textsuperscript{th} quantum), we see a steady increase in data complexity across all complexity measures for all the datasets except for the Bug-delay dataset. We see that all four complexity measures (i.e., Fischer's discriminant ratio, linear separability, mixture identifiability, and nonlinearity) are very high for the data points in the noisy area compared to the data points in the extremes, and the inclusion of such complex data makes it very hard for the classifiers to perform well. The steady increase in data complexity as we move across the quantum can be attributed to the steady increase of the discretization noise in the dataset as we move from the 1\textsuperscript{st} quantum to the 5\textsuperscript{th} quantum (the extremes to the noisy area). Therefore, when the discretization noise (the data points in the noisy area with high complexity) is discarded, the performance of some of the classifiers increases.

The lower complexity in the 2\textsuperscript{nd} quantum for the Bug-delay dataset does not impact our findings. It is due to the way the dataset is split, the BoxCox transformation aims to spread the dependent variable sufficiently so that the class-wise binning yields data in all quantum. But for the Bug-delay dataset, when we split the data into quantum, we observe that the 2\textsuperscript{nd} quantum has data points that only belong to ``class 2'' and not ``class 1'' because the quantum 2 for the Bug-delay dataset contains only data points belonging to ``class 2'', its complexity is very low, which is reflected in Figure~\ref{fig:disc2_1}. But this phenomenon has no bearing on our findings that the quantum containing high volumes of discretization noise (q5) is more complex than the quantum containing extremes data (q1) and thereby discarding them sometimes improves the performance of the classifiers. 

Hence, the presence of a high volume of discretization noise in the noisy area increases the data complexity, which in turn results in the decreased performance of a classifier that is trained with discretization noise, despite containing useful information. Therefore, in some cases, the performance of a classifier benefits from discarding the data points with from the noisy area. \label{sec:disc2}
\section{Guidelines for Using Our Framework}~\label{sec:guideline}

We explain in detail our framework in Section~\ref{sec:framework} and demonstrate how it is used to study the impact of discretization noise on the performance and interpretation of a classifier in Section~\ref{sec:rqs}. Furthermore this section, we provide practical guidelines on how to use our framework and the best practices to follow.

\begin{figure*}
    \includegraphics[width=\linewidth]{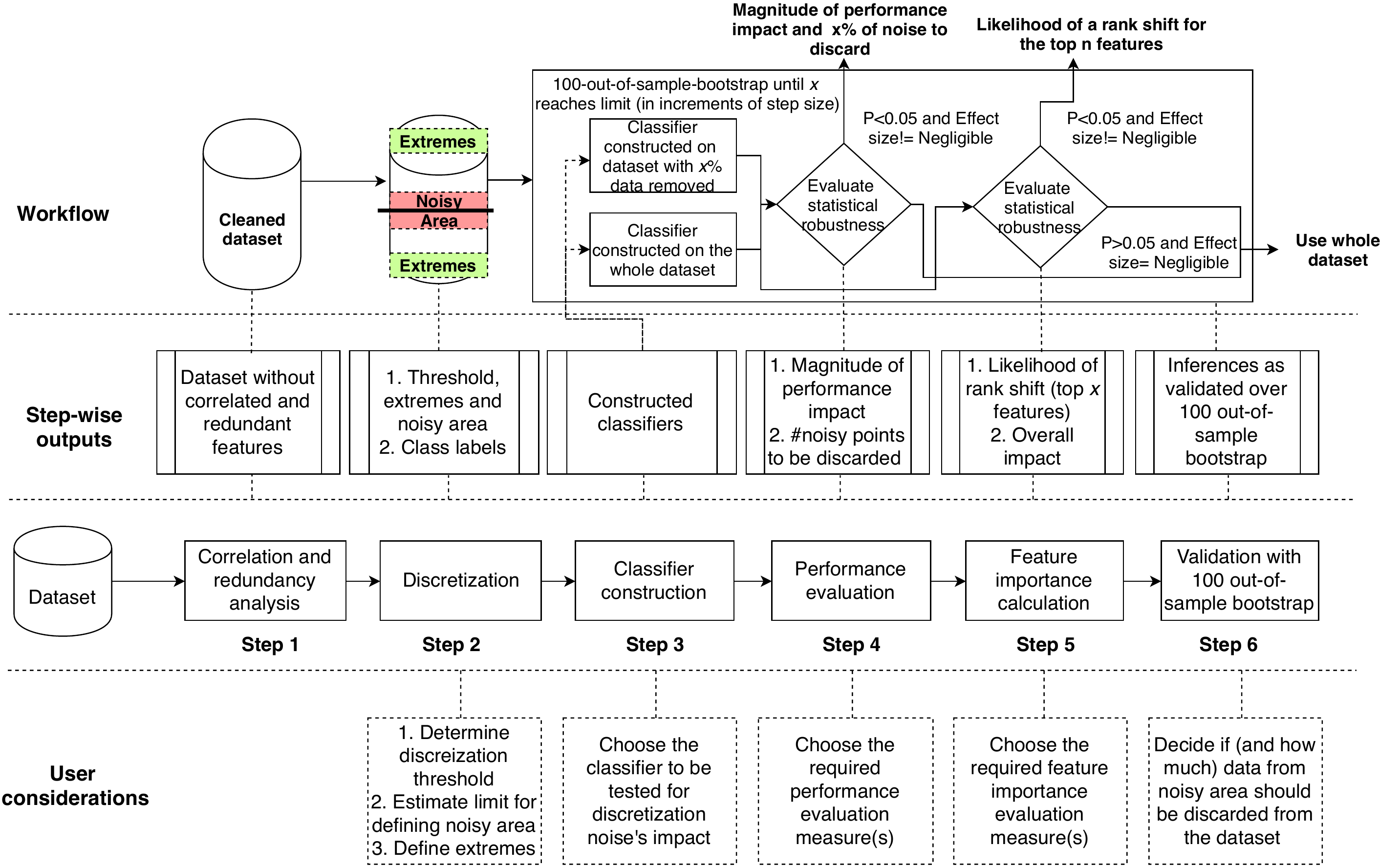}

    \caption{User considerations and workflow that are associated with each of the steps of our framework.}

    \label{fig:guidelines}
\end{figure*}




Figure~\ref{fig:guidelines} shows the involved steps, step-wise outputs, user considerations at each step and the overall workflow of our framework. A user can follow the steps one by one when they are given a dataset to study. 


\subsection{Performance impact estimation} A classifier constructed with increasing amounts of discretization noise being removed is constantly compared against the classifier constructed on the whole dataset with a Wilcoxon signed-rank test and a Cohen's effect size test as outlined in Section~\ref{sec:rq1}. If for the chosen performance measure, the impact is statistically significant with non-negligible effect size, then the amount of noisy points to be discarded and the magnitude of the performance impact due to the discretization noise is reported to the user. If the discretization noise does not impact the chosen performance measure then our framework would output 0 (suggesting no data needs to discarded) and recommend the use of the whole dataset as outlined in the workflow of Figure~\ref{fig:guidelines}.

However, the choice of the performance measure to focus on and how much of an improvement/impact that one should consider actionable depends entirely on the context. For instance, in a dataset of 100 data points, if 90 data points belong to ``class 1'' and 10 data points belong to ``class 2'', then accuracy (w.r.t ``class 1'') would be 90\% even if the classifier always predicts ``class 1'' for all the 100 data points. Therefore, other balanced performance measures like AUC might be required. 


\subsection{Interpretation impact estimation} The derived feature importance ranks of the classifiers constructed on datasets with varying amounts of discretization noise being removed is computed. These computed ranks are compared to see if the derived feature importance ranks change between the classifiers that are constructed on the whole dataset and the ones that are constructed on the dataset without discretization noise (please see Section~\ref{sec:rq2} and Figure~\ref{fig:guidelines}). Our framework then checks if the differences of the derived feature importance ranks between the classifiers that are trained on the dataset with and without the discretization noise are statistically significant. If they are, our framework also calculates the likelihood of rank shifts for the top n features (between the classifier trained on the whole dataset and the data with the framework recommended amount of data from noisy area removed). In summary, our framework reports if there is an impact of discretization noise on the overall interpretation and the likelihood of rank shifts in for the top n features to the user.



\subsection{Best practices}
In this section, we recommend the key best practices for others to follow when they are discretizing the data using an artificial threshold. From Section~\ref{sec:rq1} we note that performance of all the classifiers is impacted across all performance measures differently and that class-specific performance measures (e.g., Precision and Recall) are more sensitive to discretization noise than others. Therefore, we recommend the following best practices for the researchers and practitioners:
\begin{enumerate}[leftmargin=*]
\item Irrespective of the choice of a classifier, if one intends to discretize the continuous dependent variable into artificial classes, one should use our framework to analyse if they should use the whole dataset or discard the discretization noise.
\item Class-specific performance measures are more sensitive to discretization noise. Therefore, instead of discarding a fixed amount of data like some of the previous studies~\cite{wang2017,Tian:2015,hassan2018studying}, we recommend the use of our framework to estimate how much discretization noise that one should discard to avoid any negative impacts on their performance measure of choice. Hence, one could avoid the unwanted loss of data.
\item If our framework reports a high likelihood of a rank shift for one of the top n features, we recommend not to trust the feature importance rank for that particular feature and seek the opinion of the domain expert. However, if our framework detects any impact in the overall interpretation along with the performance, then we recommend the use of the interpretation of the best performing model.
\end{enumerate}


\section{Threats to Validity}\label{sec:threat}
\noindent\textbf{External Validity} Many of the prior studies highlight that different classifiers have different performance on the same data~\cite{rajbahadur2017impact,Ghotra:2015:RIC:2818754.2818850}. So the choice of classifiers might impact the findings of our study, as we only use four classifiers (RFCM, LR, CART, and KNN) in our analysis. However, the chosen classifiers represent a diverse range of families: statistical family, nearest neighbor family, Decision tree family and ensemble family, i.e., 4/6 of the common classifier families as outlined by Lessmann~\textit{\etal}~\cite{lessmann2008benchmarking}. We left out representative approaches from the neural networks family and the support vector machine family. We did so, as classifiers from these families typically do not have a default feature importance measure. 

\noindent\textbf{Construct Validity} Threats to construct validity pertains to the suitability of the measures that are used in our study. In our study, we study the impact of discretization noise that is generated in the dataset by using three different discretization thresholds as mentioned in Section~\ref{sec:framework} and the results might vary when another threshold is used. However, the three chosen discretization threshold computation methods (MT, CT, RTT) discretize the dependent variables differently and represent the most common ways of unsupervised discretization of the dependent variable. Our framework enables others to explore other discretization thresholds in a systematic manner.


Another construct validity in our study is the choice of the $limit$ parameter for deciding the size of the noisy area in each dataset. We used the limit values that are generated by our automated noisy area estimation algorithm as given in Section~\ref{sec:framework}. Though such an algorithm estimates the noisy area quantitatively based on a complexity measure, it might not consider the inherent dataset characteristics and bias. We acknowledge that it could be a potential threat and we urge researchers to explore various limits, or with limits that are established by domain experts and ratify our findings. Future studies should use our framework to test different values for the limit.

Finally, in this section, we wish to reiterate to the readers that our framework enables the researchers and practitioners to fiddle with any components and try a variety of combinations. We only define the needed analysis that is to be done, so that the drawn observations are valid.

\section{Conclusion}\label{sec:conclusion}


In this paper, we propose a framework to systematically and rigorously analyse the impact of discretization noise on the performance and interpretation of a classifier within the context of their own domain. We perform a case study on a variety of software engineering datasets and we find that:
\begin{enumerate}

\item Discretization noise impacts the different performance measures of classifiers differently across the different datasets. We observe that discretization noise leads to an up to 139\% performance differences across various performance measures across all the studied classifiers. Hence it is very important for researchers and practitioners to use our framework to analyse the impact of discretization noise on the classifier's for before either including or discarding it in their analysis. 
\item When discretization noise negatively impacts the performance of a classifier, our framework provides a systematic and statistically robust way to estimate exactly how much data should be discarded to avoid discretization noise without incurring unwarranted data loss.
\item Though discretization noise impacts the overall derived feature importance ranks of a classifier, it does not impact the derived feature importance ranks of the top 3 ranks for our case studies. Our framework provides a case by case guidance for others who wish to explore its use for their own case studies.
\end{enumerate}

\noindent\textbf{R package and User Guideline:} We provide an R package to enable others to use our framework to analyse the impact of discretization noise. Furthermore, we provide a user guideline, a step-by-step walkthrough and the best practices of using our framework in Section~\ref{sec:guideline}.

\ifCLASSOPTIONcaptionsoff
  \newpage
\fi



%

\bibliographystyle{IEEEtranS}
\bibliography{main}

%
\vspace{-1.0cm}
\begin{IEEEbiography}[\vspace{-0.25in}{\includegraphics[width=0.8in,height=1in,clip,keepaspectratio]{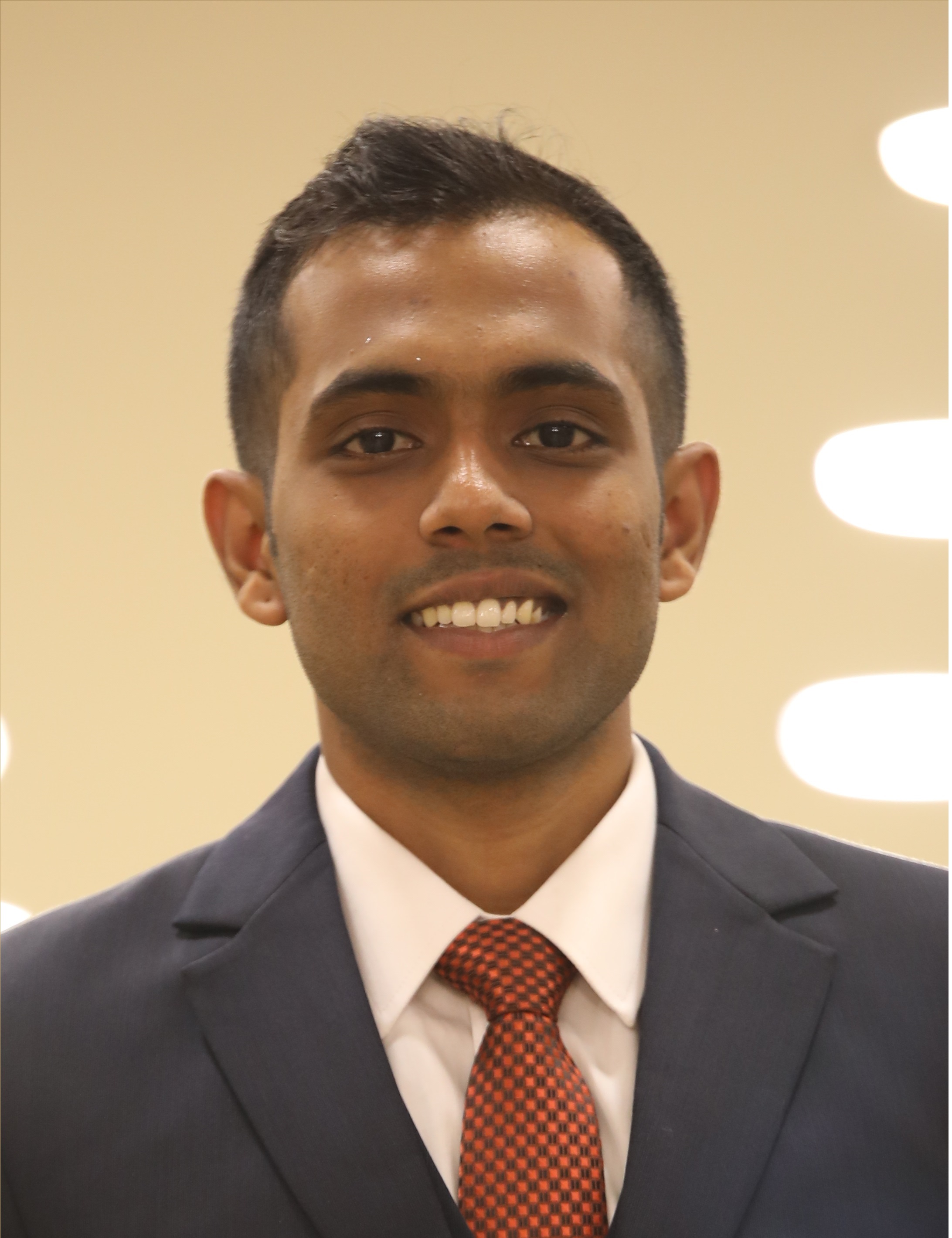}}]{Gopi Krishnan Rajbahadur}
is currently a Ph.D. student in the Software Analysis and Intelligence Lab (SAIL) at Queen's University, Canada. He received his BE in computer Science and Engineering from SKR Engineering college, Anna University, India. He also spent close to five years working as a data scientist in various software corporations in both India and Canada. His research interests include interpretable machine learning, mining software repositories and safe data science. More information at: http://gopikrishnanrajbahadur.me/
\end{IEEEbiography}
\vspace{-0.60in}

\begin{IEEEbiography}[\vspace{-0.25in}{\includegraphics[width=0.8in,height=1in,clip,keepaspectratio]{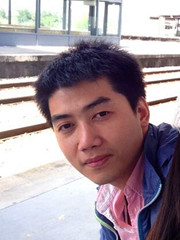}}]{Shaowei Wang}
is a postdoctoral fellow in the Software Analysis and Intelligence Lab (SAIL) at Queen’s University, Canada. He obtained his PhD from Singapore Management University, and BSc from Zhejiang University. His research interests include code mining and recommendation, software maintenance, developer forum analysis, and mining software repositories. More information at:https://sites.google.com/site/wswshaoweiwang/

\end{IEEEbiography}
\vspace{-1.9cm}

\begin{IEEEbiography}[\vspace{-0.25in}{\includegraphics[width=0.8in,height=1in,clip,keepaspectratio]{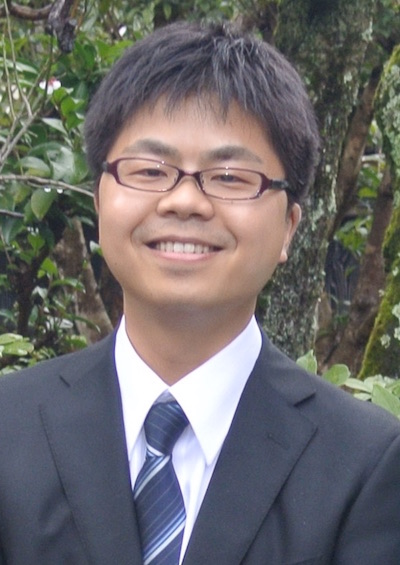}}]{Yasutaka Kamei} is an assistant professor at Kyushu University in Japan. He received his BE degree in informatics from Kansai University, and PhD degree in information science from the Nara Institute of Science and Technology. He was a research fellow of the JSPS (PD) from July 2009 to March 2010. From April 2010 to March 2011, he was a postdoctoral fellow at Queen’s University in Canada. His research interests include empirical software engineering, open source software engineering, and mining software repositories (MSR). He is a member of the IEEE More information at http://posl.ait.kyushu-u.ac.jp/~kamei/
\end{IEEEbiography}
\vspace{-1.3cm}

\begin{IEEEbiography}[\vspace{-0.25in}{\includegraphics[width=0.8in,height=1in,clip,keepaspectratio]{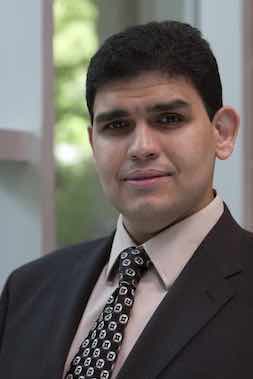}}]{Ahmed E. Hassan} is an IEEE fellow and member, ACM influential educator, an NSERC Steacie Fellow, a Canada Research Chair (CRC) in Software Analytics, and the NSERC/BlackBerry Software Engineering Chair at the School of Computing at Queen’s University, Canada. His industrial experience includes helping architect the Blackberry wireless platform, and working for IBM
Research at the Almaden Research Lab and the Computer Research Lab at Nortel Networks. His research interests include mining software repositories, empirical software engineering, load testing, and log mining. Dr. Hassan serves on the editorial board of the IEEE Transactions on Software Engineering, the Journal of Empirical Software Engineering, and PeerJ Computer Science. He spearheaded the organization and creation of the Mining Software Repositories (MSR) conference and its research community. More information at https://sail.cs.queensu.ca/
\end{IEEEbiography}




\end{document}